\documentclass[10pt, journal, twocolumn]{IEEEtran}

\usepackage[dvips]{graphicx}
\usepackage{amsmath,amssymb}
\usepackage{amsfonts}
\usepackage{mathabx}
\usepackage[table,usenames,dvipsnames]{xcolor}
\usepackage{bbold}
\usepackage[normalem]{ulem}
\usepackage{soul}

\begin{document}

\newcommand{\Eq}[1]{Eq.~\eqref{#1}}
\newcommand{\Eqs}[1]{Eqs.~\eqref{#1}}
\newcommand{\flqu}{\frac{\Phi_0}{2\pi}}
\newcommand{\phiBB}{\phi^{B}}
\newcommand{\phiAA}{\phi^{A}}
\newcommand{\phiCC}{\phi^{C}}
\newcommand{\IJab}{\hat I_c}
\newcommand{\CJab}{\hat C_J}
\newcommand{\Lhat}{\hat L}
\newcommand{\La}{\hat{L}^A}
\newcommand{\Lb}{\hat{L}^B}
\newcommand{\Lc}{\hat{L}^C}
\newcommand{\IJaa}{I_c^{A}}
\newcommand{\CJaa}{C_J^{A}}
\newcommand{\IJbb}{I_c^{B}}
\newcommand{\CJbb}{C_J^{B}}
\newcommand{\IJcc}{I_c^{C}}
\newcommand{\CJcc}{C_J^{C}}

\newcommand{\myclock}{\text{clk}}
\newcommand{\Efl}{E_{\text{fl}}}

\title{Reversible Fluxon Logic with optimized CNOT gate components}
\author{K.D.~Osborn and W.~Wustmann%
\thanks{K.D.~Osborn and W.~Wustmann are with The Laboratory for Physical Sciences at the University of Maryland, College Park, MD 20740, USA. K.D.O. (osborn@lps.umd.edu) is additionally with the Joint Quantum Institute and Quantum Material Center at the University of Maryland, College Park.}}
\maketitle

\begin{abstract}
Reversible logic gates were previously implemented in superconducting circuits 
as {\em adiabatic-reversible} gates, 
which are powered with a sufficiently slow clock. 
In contrast, we are studying {\em ballistic-reversible} gates, 
where fluxons serve to both encode the information and power the gates.
No power is applied to the gate apart from the energy of the input fluxons,
and the two possible flux polarities represent the bit states.
Undamped long Josephson junctions (LJJs), 
where fluxons move at practically constant speed from inertia, 
form the input and output channels of the gates.
LJJs are connected in the gates by circuit interfaces, which are designed 
to allow the ballistic scattering from input to output fluxon states, 
using the temporary excitation of a localized mode. 
The duration of the resonant scattering determines the operation time of the gate, 
approximately a few Josephson plasma periods. 
Due to the coherent conversions between fluxon and localized modes
the ballistic gates can be very efficient:
in our simulations only a few percent of the fluxon's energy are 
dissipated in the gate operation.
Ballistic-reversible gates can be combined with other, non-ballistic gate circuits 
to extend the range of gate functionalities.
Here we describe how the CNOT can be built as a structure that includes
the IDSN (Identity-else-Same-gives-NOT) and Store-and-Launch (SNL) gates. 
The IDSN is a 2-bit ballistic gate, which we describe and analyze in terms
of equivalent 1-bit circuits.
The SNL is a clocking gate, that allows the storage of a bit
and the clocked launch of a fluxon on a bit-state dependent output path. 
In the CNOT the SNL gates provide the necessary routing and 
fluxon synchronization for the input to the IDSN gate.\\
\end{abstract} 

\begin{IEEEkeywords}
Ballistic signaling, 
fluxons, 
power efficient, 
reversible computing,  
superconducting logic circuits. 
\end{IEEEkeywords}

\section{Introduction}

Semiconducting logic greatly benefits from the dramatic scaling down 
of transistor gate and interconnect dimensions.
Nowadays, however, the benefits from further scaling are nearly exhausted.
This has direct implications for the characteristics of logic gates
such as a gate's energy cost:
one of the bit states is stored as a voltage state on a capacitor,
and the stored bit energy scales with the dimensions of the latter.
In bit switching, which here amounts to discharging the capacitor,
the entire energy of the stored bit state is dissipated. 
This energy cost is still much higher than 
the theoretical minimum energy cost, $\log(2) k_B T$, 
incurred for every bit erasure in irreversible logic gates.
In reversible logic gates, on the other hand, no bit erasure takes place.
The ensuing absence of a fundamental lower bound of the energy cost
motivated Bennett to develop 
a mathematical model for a reversible computer \cite{Bennett1973}. 
Later, Likharev described how classical reversible computing could be achieved 
in a superconducting circuit, using an external (clock) drive that 
adiabatically modulates the circuit potential \cite{Lik1982}. 
The energy for bit switching is here proportional
to the inverse of the gate time, and thus can theoretically be 
lowered indefinitely.
More recently, such {\em adiabatic-reversible} gates have been realized
with superconducting technology, 
in circuits named the Quantum Flux Parametron (QFP) \cite{QFPgate} 
and the N-SQUID \cite{RenSem2011}.

A physically different approach for energy-conservative computing
is based on {\em ballistic-reversible} gates.
In the classic model for these gates \cite{FredTof1982},
logic operations are defined by the scattering of billiard balls.
The gates are powered by the inertia of the input states (billiard balls) alone, 
in contrast to the external drive (clock) power used in the adiabatic model.
Ballistic-reversible gates have been studied using optical
solitons as information carriers \cite{SocETAL1992, SchOre2005}, 
however, these optical ``particles'' have been more thoroughly 
investigated for high-speed and long-distance communication \cite{FerETAL2005}.

We have previously proposed Reversible Fluxon Logic (RFL), 
which is designed to realize ballistic-reversible gates by 
the scattering of fluxons in special gate circuits \cite{WusOsb2020}. 
The ballistic gates are unpowered other than the energy of the bit-representing fluxons. 
Also, the number of fluxons scattered in such a gate is conserved, and their energy is nearly conserved. 
However, the ballistic gates of RFL are not simply a realization of the billiard-ball model with fluxons, 
but have two main distinct features: 
(i) flux-polarity changes determine bit-switching in the gates instead of path changes, and 
(ii) the scattering processes which define the gate operation are resonant. 
The latter feature makes the gates both energy-efficient and fast, 
with gate duration set by a few natural JJ oscillation cycles.

\begin{table}[tb]\centering
\caption{ 
Comparison of 
different digital logic technologies.
In contrast to irreversible types (CMOS, Irreversible SFQ),
in reversible logic types (adiabatic-rev. and ballistic rev. SFQ), 
the energy cost can be small relative to the stored bit energy.
}
\renewcommand\arraystretch{1.5}
\tabcolsep=0.15cm
\scalebox{0.8}{\sf 
\begin{tabular}{|p{1.2cm}|p{1.6cm} p{1.6cm} p{2.2cm} p{2.8cm}|}
\hline
 & CMOS & Irreversible SFQ & Adiabatic-rev. SFQ & Ballistic-rev. RFL \\
\hline
bit states\newline 1 \& 0 & voltage state\newline \& null & flux state\newline (SFQ) \& null & equiv.~circulating\newline current states\newline (e.g. CW \& CCW)
& fluxon polarity $\pm 1$\newline (topological charge) \\
\rowcolor[gray]{0.8}
stored bit\newline energy & $CV^2/2$\newline charging\newline energy & $\sim I_c \Phi_0$\newline JJ-switching\newline energy &  $\lesssim I_c \Phi_0$\newline (time-dependent) & 
$\Efl=$\newline$8E_0\big/\sqrt{1-v^2/c^2}$\newline fluxon energy  \\
energy cost & $CV^2/2$ per\newline bit switching & $I_c \Phi_0$ per\newline switching JJ & $\ll I_c \Phi_0$ & $\ll \Efl$ \\
\rowcolor[gray]{0.8}
power source & voltage bias\newline $V$ & current bias\newline $I \lesssim I_c$ & (multi-phase)\newline current bias  & excess bit energy (e.g.\newline kinetic fl. energy) \\
\hline
\end{tabular}
}
 \label{table:compare_logic}
\end{table}

Table \ref{table:compare_logic} compares different logic types
in terms of their bit states, stored bit energy, energy cost, and power source.
In CMOS (first column), the stored bit energy is given by the charging energy
of a capacitance $C$ held at a source voltage $V$. 
The capacitor is discharged during bit switching, and the entire charging energy 
is lost in the process.
Similarly, in irreversible SFQ logic (e.g., RSFQ, ERSFQ, RQL)  (second column)
one bit state is represented by a single flux quantum (SFQ) stored at an energy
$\sim I_c \Phi_0$. This is approximately equal to the energy cost of bit switching,
which occurs under a $2\pi$ phase slip of a damped JJ with critical current $I_c$.
For the bit states to be distinguishable,
they must be separable by a potential barrier that is large compared to the 
energy of thermal fluctuations $k_B T$, 
thus requiring $I_c \Phi_0 \gg k_B T$.

In contrast, reversible logic gates preserve a significant part of the stored bit energy. 
One approach to this in superconducting circuits are adiabatic-reversible gates (third column), 
where the bit-defining potential is slowly modulated by a clock. 
The two bit states are usually represented by circuit states of equal energy. 
During the transition from one bit state to the other
the energy dissipated in the circuit scales inversely with the clock period \cite{Lik1982} 
and thereby can be made arbitrarily small. 

With RFL (fourth column)
we follow the alternative approach of ballistic-reversible gates, 
which is based on the undamped motion of fluxons in long Josephson junctions (LJJs).
A fluxon in an LJJ contains the flux of one SFQ, $\pm \Phi_0$,
and according to the sign (polarity) it is designated as either a fluxon or an antifluxon.
The two flux polarities 
are used to represent the two bit states.
Both bit states have the same stored bit energy, given here by the fluxon energy $\Efl$
which is composed of the rest (potential) energy 
$8 E_0 = 8 I_c \Phi_0 \lambda_J/(2\pi a)$ and kinetic energy. 
This stored bit energy can be adjusted for the application, 
similar to other superconducting SFQ technologies.
Ballistic RFL gates make use of the scattering of fluxons at 
special circuit interfaces between LJJs.
The fluxon's rest energy and the fluxon number is conserved in the scattering.
Moreover, as our simulations of the classical equations of motion of the RFL 
gate circuits show, the total fluxon energy is also conserved to a large extent. 
The remainder is lost to small-amplitude plasma waves
generated by the fluxon at the gate.

Ballistic RFL gates developed so far have no internal state memory,
in contrast to proposed
asynchronous reversible gates \cite{Frank2017, FrankETAL2019, FrankETAL2019_ISEC}.
That is why, ballistic gates for two or more input bits, 
such as the 2-bit NSWAP \cite{WusOsb2020} and IDSN (Identity-else-Same-gives-NOT) 
gates \cite{OsbWus2018},
require a synchronous arrival of the input fluxons.
In order to reliably use these gates as part of a larger circuit, 
we have therefore developed a gate for the purpose of clocking and synchronization.
According to its operation it is named 
a Store-and-Launch (SNL) gate \cite{OsbWus2018}. 
The gate stores the bit state of the incoming data fluxon as a static circulating current. 
Later, upon arrival of a low-energy clock fluxon, the state is launched 
as a fluxon with the same polarity as the original. 
The clock fluxon is the sole power source of the launch and is annihilated in the process. 
Part of its energy goes to the launched data fluxon
which then may have larger energy than the (slowed-down) input data fluxon.
The SNL can be designed with a clock fluxon, which has only a fraction of the data fluxon's energy, and is therefore energy efficient by irreversible logic standards.
Moreover, with SNL-clock fluxons as the main energy cost,
while ballistic gates and LJJs as transmission lines are unpowered, 
RFL can avoid problems associated with the DC-biasing of JJs
in irreversible SFQ logic \cite{TerETAL2003, KanKap2003, MukhETAL2013}.

The launch direction of the SNL is bit-state dependent: 
bit state 0 (fluxon) and bit state 1 (antifluxon) each have their designated output LJJ. 
This bit-state dependent routing adds an important resource to RFL logic.
It allows to combine basic RFL components in a way that achieves 
a complex reversible gate operation, which for symmetry reasons 
cannot be achieved directly with a single ballistic RFL gate.
As an example, we had proposed in Ref.~\cite{OsbWus2018} 
to implement a CNOT with a structure 
that is composed of several ballistic gates (including two IDSN gates) 
together with two SNL gates. 

RFL was originally introduced in Ref.~\cite{WusOsb2020}, 
where we studied fundamental ballistic gates of RFL 
through numerical simulation and also analytically with a collective coordinate model.
In Ref.~\cite{OsbWus2018} we had extended the scope of RFL, by introducing concepts 
of the clocking gate SNL, the ballistic IDSN and the composite CNOT. 
These gates were however not studied in great detail and had not been optimized. 
Here we provide the missing details and extend the original study:
for the IDSN we provide optimized parameters and margins, and also 
introduce an analytic model from which we derive parameter equivalences with 1-bit gates.
We now introduce and study an SNL with two input LJJs, 
as is required in the CNOT application, 
while initially only a 1-input SNL version had been described~\cite{OsbWus2018}.

This article is organized as follows: 
Firstly we summarize in Sec.~\ref{sec:RFL} the operation of ballistic RFL gates and explain the resonant scattering dynamics in detail using the example
of the simulated 1-bit NOT gate, Sec.~\ref{sec:NOT}.
The optimized 2-bit IDSN gate is described in Sec.~\ref{sec:IDSN},
where we show the simulated dynamics and analytically map the gate 
to equivalent 1-bit gates.
Sec.~\ref{sec:SNL} presents the 2-input SNL gate 
with details of the dynamics and a comparison to 
an earlier 1-input SNL \cite{OsbWus2018}. 
In Sec.~\ref{sec:CNOT} we describe the CNOT gate which is composed of the 
gates from the preceding sections.

\section{Short summary of ballistic RFL gates}\label{sec:RFL}

\begin{figure}[tb]\centering
\includegraphics[width=0.97\columnwidth]{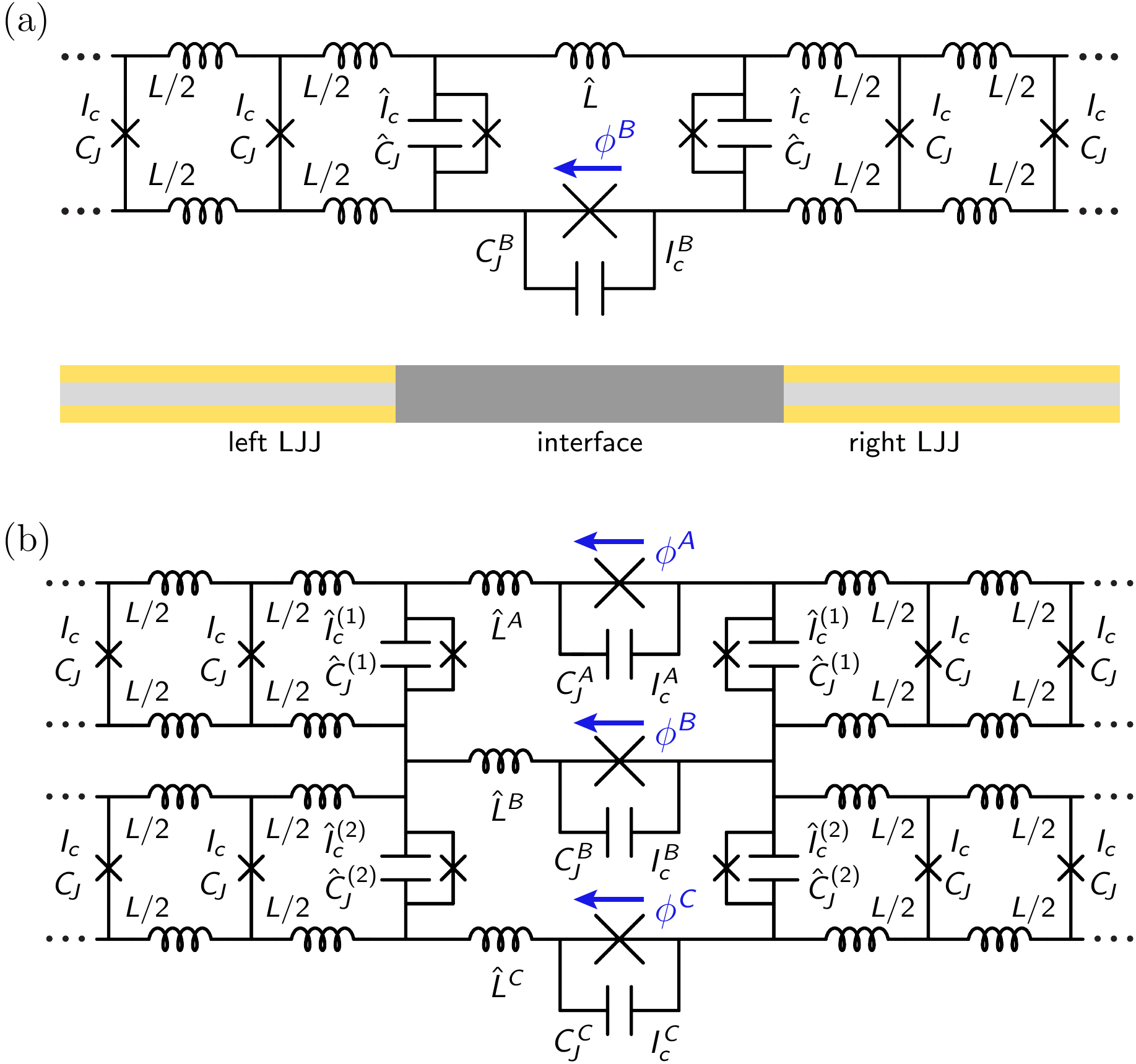}
\caption{%
RFL gate structures: (a) Schematic for a ballistic 1-bit gate,
consisting of two LJJs connected by an interface cell with three capacitance-shunted JJs:
the left and right `termination JJs' with $(\CJab,\IJab)$,
and the `rail JJ' with $(\CJbb, \IJbb)$.
For a ballistic RFL gate, parameters have to be set such that an
incident free fluxon undergoes the desired type of forward-scattering 
from one LJJ to the other. 
These resonant dynamics are enabled by specific large values of the shunt capacitances 
$\CJbb, \CJab$. 
The sketch under the schematic illustrates the essential structure of the gates; 
dynamics in the discrete LJJs is similar to the continuous limit. 
(b) Schematic for ballistic  2-bit gates consisting of two input and two output LJJs, 
and a circuit interface with 7 JJs. 
These allow forward-scattering of fluxons similar to 1-bit ballistic gates, but
now also with input-dependent (conditional) polarity changes.
In efficient ballistic gates the interface's rail inductance(s) are small, $\Lhat \ll L$, 
and this prevents the interface cell(s) from storing a flux quantum.
}
 \label{fig:gateschematics}
\end{figure}

Long Josephson junctions (LJJs) are key structures in RFL. 
We design RFL circuits with discrete forms of LJJs,
where an array of identical Josephson junctions (JJs) are connected by two 
inductor ``rails''. 
In Fig.~\ref{fig:gateschematics}(a) two such discrete LJJs form the left and right parts of the circuit. 
The JJs have capacitance and critical current of $(C_J, I_c)$, 
and each unit cell of length $a$ has the inductance $L$.
The circuit parameters set the Josephson plasma frequency $\omega_J = 2\pi \nu_J = \sqrt{2\pi I_c/(\Phi_0 C_J)}$, and the Josephson penetration depth, 
$\lambda_J = a \sqrt{\Phi_0/(2\pi L I_c)}$. 
The latter determines the length scale of phase gradients in the (discrete) LJJ, 
such as the width of a fluxon or edge states at the LJJ boundaries.

A fluxon in an LJJ is described by the soliton solution 
$\phi(x,t) = 4 \arctan\left(\exp\left(\mp (x - v t)/\sqrt{1-v^2/c^2}\right)\right)$
of the Sine-Gordon equation 
$\partial_{tt} \phi - c^2 \partial_{xx} \phi + \omega_J^2 \sin\phi = 0$ 
for the superconducting phase field $\phi(x)$.
According to that solution, the fluxon behaves like a relativistic particle, 
moving with constant speed $v \leq c$ below the upper velocity bound 
$c=\omega_J \lambda_J$.
The energy of the moving fluxon is increased 
by a factor
\begin{equation}\label{eq:Efl_vdependence}
 \Efl(v)/\Efl(0) = \left(1-(v/c)^2\right)^{-1/2}
\end{equation}
relative to its rest energy $\Efl(v=0) = 8 E_0$,
where $E_0 = I_c \Phi_0 \lambda_J/(2\pi a)$.

In a discrete LJJ, the discreteness introduces a damping of the fluxon motion,
compared to its motion at constant speed in a continuous LJJ.
The strength of this perturbation is determined by the relative discreteness $a/\lambda_J$.
In our circuit simulations, we choose $a/\lambda_J \simeq 1/3$ to be sufficiently small, 
such that the loss in speed (and energy) is 
negligible even when the fluxon moves over hundreds of unit cells.

The bit-switching mechanism of the logic requires a method to invert the fluxon polarity.
The 1-bit NOT gate, as the fundamental realization for bit-switching,
is implemented in a circuit as shown in Fig.~\ref{fig:gateschematics}(a),
where an input LJJ and output LJJ are connected by a circuit interface.
The designation as input and output LJJs can in principle be reversed 
since the circuit has left--right symmetry.
The interface consists of at least three capacitively shunted JJs (CSJJs).
Each LJJ is terminated by a CSJJ with parameters $(\CJab,\IJab)$
between its inductor rails, 
and we call these the left and right `termination-JJs' of the interface.
The two rails of one LJJ are connected to those of the other, 
where one connection (shown between the lower rails)
is formed by a CSJJ with parameters $(\CJbb,\IJbb)$, 
and we call it the `rail-JJ' of the interface.
The other connection (shown between the upper rails) 
is made with an inductor $\Lhat$ that is typically small ($\Lhat \ll L$)
and the interface cell thus cannot store a flux quantum.

Depending on the interface parameters, such a circuit interface 
has been found to enable forward-scattering of an incoming fluxon 
from one LJJ to the other \cite{WusOsb2020}. 
In these processes the incident fluxon breaks at the interface into two parts, 
where its characteristic phase and current distributions become discontinuous. 
Its energy is transferred to excitations of the interface JJs and evanescent 
fields in the LJJs close to the interface. 
These localized excitations undergo a short coherent oscillation 
before a new fluxon forms in the other LJJ. 
Importantly, for certain interface parameters the polarity of the newly created 
fluxon is opposite to that of the original fluxon, and thus realizes a NOT gate.
We note that polarity inversion is not possible for a fluxon 
within the bulk of a planar LJJ due to its topological nature 
(other than by a wiring crossover for a half-twist in an LJJ, not used here). 
In contrast, in the NOT gate the polarity inversion is possible
due to the interface's rail-JJ, which opens the otherwise flux-impermeable LJJ 
rails and allows for a large difference of $4\pi$ 
building up between the left and right termination-JJ phases. 
This phase change 
is dissipationless, unlike phase switching in the resistively shunted JJs 
of SFQ logic gates or in overdriven unshunted JJs. 
In subsection~\ref{sec:NOT} below we describe the NOT gate operation in more detail.

For the purpose of ballistic gates, the parameters of the interface cell are set 
such that the fluxon scattering is resonant. 
By resonance we mean that the energy transfer from the moving fluxon to the localized excitation 
and again to a moving fluxon happens coherently. 
Considering the highly nonlinear nature of the fluxons and other involved modes, no obvious 
and unambiguous resonance criterion in the sense of matching frequencies exists.
At resonance, the gate is particularly efficient 
since energy loss through plasma waves generated at the interface is minimized. 
Correspondingly, the energy-efficiency of the gate (ratio of fluxon energy before and after the scattering) 
assumes a local maximum with respect to most interface parameters, cf.~Fig.~9 in Ref.~\cite{WusOsb2020}.
For example, a resonant NOT gate requires relatively large shunt capacitances 
of the three interface JJs \cite{WusOsb2020, Liuqi2019}.

Different types of scattering resonances can be observed 
at different points in parameter space, 
each within a finite (but wide) range around an efficiency maximum.
They can be classified according to characteristics such as 
(i) the number and duration of the interface oscillation cycles 
(comparable to the `bounces' in LJJs with point defects \cite{GooHab2007}).
Furthermore, (ii) a resonance may preserve or invert the polarity of the incoming fluxon,
and (iii) the new fluxon may be created in the output LJJ (forward-scattering) 
or in the input LJJ (backward-scattering).
For a given set of interface parameters, the resonant scattering typically is observed 
within a range of fluxon input velocities $v$.  
In the adiabatic limit, i.e. for very small input velocity, $v \ll c$,  
the flux quantum will be lost in the gate. 
In contrast, in adiabatic-reversible logic the clock can 
be slowed in principle to an arbitrarily small frequency.

Similar to the 1-bit NOT gate, a ballistic 1-bit Identity (ID) gate 
can be implemented with the structure in Fig.~\ref{fig:gateschematics}(a),
using a specific set of interface parameters.  
The fluxon induces at the ID gate interface a different resonance compared to the 
NOT gate resonance, and from it a new fluxon is created in the output LJJ 
which has the same polarity as the input fluxon. 
This resonant ID gate can in principle be parametrically 
tuned into a NOT gate, 
since both fundamental 1-bit gates are based on the same circuit-interface structure. 
This is an advantage over a trivial ID operation in form of non-resonant 
transmission, e.g.~through a regular LJJ cell,
and informs the construction of more complex gates 
from the fundamental 1-bit gates.
We have for example designed ballistic 2-bit gates based on the circuit structure 
shown in Fig.~\ref{fig:gateschematics}(b). 
It has two input and two output LJJs connected by a circuit interface with 7 CSJJs. 
Interface parameters are set such that a fluxon coming in on the upper (lower) input LJJ is scattered forward to the upper (lower) output LJJ.
In these 2-bit gates the polarity inversion is conditional:
depending on the interface parameters and on the presence and polarity of a synchronized input fluxon on the other LJJ, the action on the bit is polarity-preserving or polarity-inverting.
So far, ballistic RFL designs exist for a 2-bit NSWAP=NOT(SWAP) logic gate \cite{WusOsb2020}, 
and for an IDSN logic gate \cite{OsbWus2018}. 
An updated version of the latter is discussed in detail in Sec.~\ref{sec:IDSN}.

The dynamics of the JJ phases in a given circuit is obtained 
through numerical integration of the classical circuit equations of motion,
with fluxon(s) moving in the input LJJ(s) taken as initial conditions.
For a desired gate operation, we identify suitable interface parameters 
by optimization of the gate efficiency.
Parameter variations around these optimized values generally show
relatively robust parameter margins \cite{WusOsb2020}, 
suggesting the feasibility of the gates in fabrication and operation.
If the interface parameters are chosen far away from the parameter range set by these margins, 
the resulting dynamics in general are not useful as an RFL gate, 
such as fluxon annihilation or reflection.
In certain limits far away from our gate margins, the fluxon dynamics at the interface is comparable to other well-studied fluxon phenomena, 
e.g. scattering of a fluxon at an LJJ end with specific boundary condition \cite{CostabileETAL1978, GhoZam1994}, 
or scattering within the LJJ at a perturbation potential.
The latter, which include the resonant chaotic scattering at point defects \cite{GooHab2007, FeiKivVaz1992} and
fluxon scattering at a qubit-induced potential \cite{AveRabSem2006}, 
do not allow polarity change. 
The interface scattering used in RFL gates is non-perturbative,
since the fluxon breaks up and loses its identity at the interface.

\subsection{Dynamics in the 1-bit NOT gate}\label{sec:NOT}

\begin{figure}[tb]
\includegraphics[width=0.97\columnwidth]{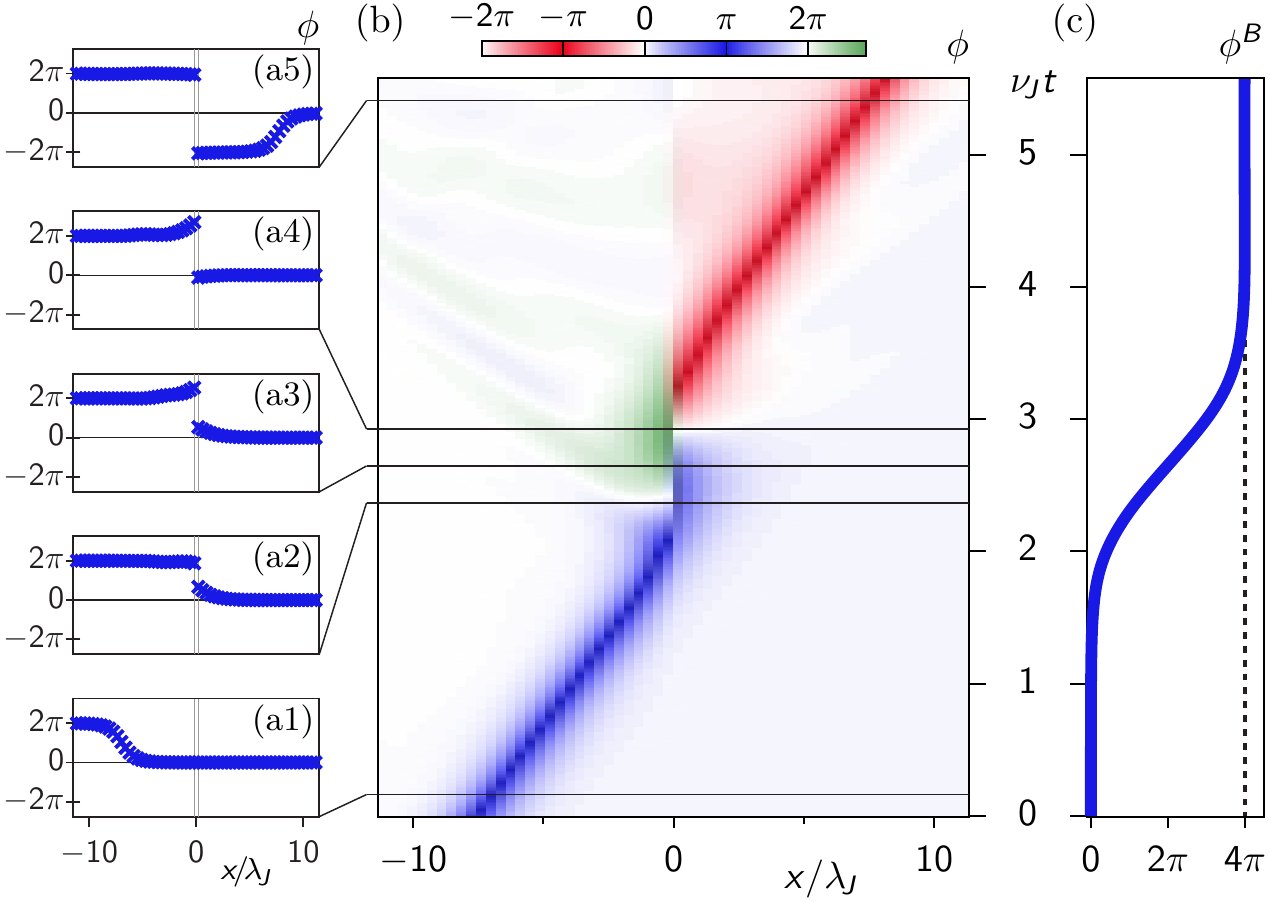}
\caption{%
RFL NOT gate dynamics. (a1-a5) JJ-phases $\phi_n$ left and right of the interface vs. position, 
at fixed times of the dynamics, 
reproduced with parameters from Ref.~\cite{WusOsb2020}: 
(a1) input fluxon moving in left LJJ towards interface;
(a2-a4) intermediate excitation of left and right interface JJ and evanescent 
phase fields left and right of the interface; 
(a5) antifluxon emitted from the interface into the right LJJ.
(b) Dynamics of JJ phases $\phi_n$ vs. time. 
The colormap, which emphasizes phases of high Josephson energy
around values $2(k+1)\pi$ ($k \in \mathbb{Z}$),
shows the center of the free fluxon (antifluxon) moving  
towards (away from) the interface at constant velocity before (after) the scattering, 
and the localized oscillation of the evanescent fields which are resonantly excited by the fluxon. 
(c) Evolution of phase $\phiBB(t)$ of the interface's rail-JJ, 
showing an adiabatic $4\pi$-change.
}
 \label{fig:NOT}
\end{figure}

Fig.~\ref{fig:NOT} shows the dynamics of a ballistic NOT gate, 
in the circuit of Fig.~\ref{fig:gateschematics}(a).
We obtain the evolution of the JJ phases from the simulation of the circuit equations of motion.
The simulation starts from an initial phase distribution corresponding to a fluxon that moves in the left LJJ ($x<0$) with velocity $\dot x = v$ towards the interface at $x=0$. 
(Note that other simulations include also a circuit structure for launching fluxons \cite{WusOsb2020}.)
The JJ phases $\phi_n$ in the left and right LJJ, 
including those of the left and right termination JJs of the interface,
are shown in the panels of subfigure (a), 
each taken at a particular time of the dynamics.
Subfigure (b) shows the evolution of $\phi_n(t)$ as a continuous function of time; 
the colormap emphasizes phases close to the values $2(k+1)\pi$ ($k \in \mathbb{Z}$)
of largest Josephson energy.
Subfigure (c) shows the evolution of the interface's rail-JJ phase $\phiBB$.

In panel (a1) one sees in the left LJJ the still undisturbed fluxon, 
with characteristic phase profile varying in position from $2\pi$ to $0$ 
(for fluxon with positive polarity).
Panels (a2)--(a4) illustrate the situation after the fluxon has broken up at the interface
and induced an excitation of the interface JJs and adjacent JJs.
This excitation has the form of exponentially localized edge states in the two LJJs.
When the fluxon breaks up (at a time just before that shown in panel (a2))
the phase profiles left and right of the interface have negative, fluxon-like slopes:
from $2\pi$ in the fluxon's wake in the left LJJ at $x \ll -\lambda_J$ 
to $\phi( x \lesssim 0 ) \approx \pi$ near the interface, 
and from $\phi(x \gtrsim 0) \approx \pi$ near the interface to $0$
in the right LJJ at $x \gg \lambda_J$. 
These phase distributions then undergo coherent amplitude swings, 
stills of which are shown in panels (a2)--(a4).
The oscillations left and right of the interface occur around $2\pi$ and $0$, 
respectively, with maximum amplitudes of $\approx \pi$, 
and with evolving phase difference between the two.
During the oscillations the LJJ phase distributions close to the interface
change their character from fluxon-like (negative slope) to antifluxon-like (positive slope).
When the slope to the right of the interface turns from negative in panel (a3) to positive in panel (a4), this may be seen as the starting point for the formation of an antifluxon. 
The phase distribution in the right LJJ develops into that of an antifluxon, varying from $-2\pi$ at the interface to $0$ in the bulk of the right LJJ. 
All along, the growing phase gap between right and left LJJ
is compensated by the likewise growing phase of the interface's rail-JJ, 
$\phiBB \approx \phi(x \gtrsim 0) - \phi(x \lesssim 0)$,
cf.~subfigure (c), 
while the small inductance $\Lhat \ll L$ in the interface cell stores negligible flux. 
By the time when $\phiBB$ has grown close to $4\pi$, 
the antifluxon is released from the interface into the right LJJ where it moves freely, 
as shown in panel (a5). 
The $4\pi$-phase change is adiabatic, 
happening on the time scale of the Josephson period, $1/\nu_J$,
in contrast to the rapid (and dissipative) $2\pi$-phase slips in RSFQ logic.

In subfigure (b) one can see the evanescent phase fields excited around 
the interface during the resonant process. 
The coherent oscillation of these edge states
is characteristic for the resonant fluxon scattering used in RFL. 
A related `bounce'-resonance can be observed in an LJJ 
when a fluxon is scattered at a point defect 
such as a locally modified critical current 
\cite{GooHab2007}.
Subfigure (b) also shows that small plasma waves are emitted into the LJJs during and after the scattering process. They carry away a fraction of the initial fluxon energy.
However, as indicated by the indistinguishable trajectory slopes of the incoming fluxon and outgoing antifluxon, this is only a minor loss: from fluxon fits before and after the scattering we see that 97\% of the fluxon energy is conserved.

\section{The IDSN gate}\label{sec:IDSN}

\begin{figure}[tb]\centering
\includegraphics[width=\columnwidth]{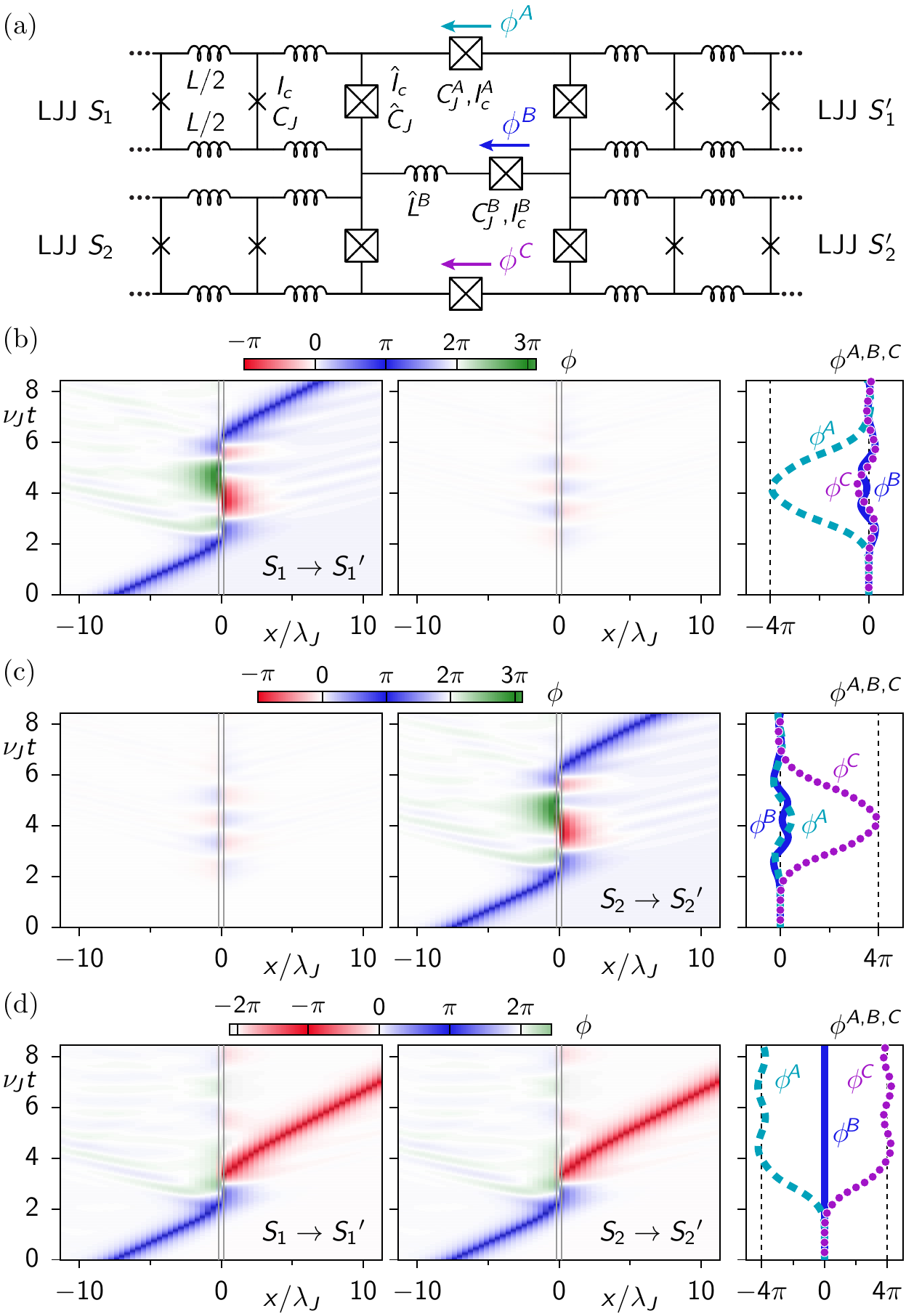}
\caption{IDSN gate circuit and dynamics.
(a) Circuit for ballistic IDSN gate. The circuit has both left-right symmetry 
and bottom-top symmetry (symmetry-related circuit elements are not labeled).
The interface JJs are shown with large symbols, which include 
large shunt-capacitances.
(b-d) IDSN dynamics for different fluxon inputs. 
The first two panels of each row show the JJ-phases $\phi_n$ as colormaps, 
one for the upper LJJs, $S_1 (x<0)$ and $S_1' (x>0)$, 
and one for the lower LJJs, LJJ $S_2 (x<0)$ and $S_2' (x>0)$.
The third panel in each row shows phases  $\phiAA,\phiBB,\phiCC$ 
of interface rail-JJs.
(b,c) A single input fluxon, which enters on (b) LJJ $S_1$ and (c) LJJ $S_2$, 
respectively, and scatters forward with preserved polarity.
(d) Two synchronized fluxons of the same polarity, 
each entering on one of LJJs $S_1$ and $S_2$, 
each scatter forward to opposite polarity state. 
The interface parameters are: 
$\CJaa/C_J = 15.0$, $\IJaa/I_c = 1.5$,
$\CJbb/C_J = 16.7$, $\IJbb/I_c = 6.9$,
$\CJab/C_J = 5.8$, $\IJab/I_c = 1.5$,
and $\Lb/L = 0.5$.
}
\label{fig:IDSN}
\end{figure}

We had recently introduced a type of ballistic 2-bit gate, 
which is named IDSN after the operations it performs:
a single input fluxon undergoes an \underline{ID} operation, 
and two synchronized fluxons of the \underline{s}ame polarity each undergo 
a \underline{N}OT operation \cite{OsbWus2018}.
Table \ref{table:IDSN_logictable} summarizes the logic action of the IDSN gate.
Note that the IDSN is a conditionally reversible gate 
\cite{Frank2017_RCarticle, Frank2017_RCarticle_extended}, 
where only certain 2-bit inputs are allowed, 
while all 1-bit inputs are allowed.
Like the previously introduced NSWAP gate, the IDSN gate is implemented 
by the resonant fluxon scattering at a gate interface of the type shown in Fig.~\ref{fig:gateschematics}(b).
The circuit has left--right symmetry across the interface, 
as well as vertical symmetry about the B-rail, i.e.~$\Lc=\La$, 
$\CJcc=\CJaa$, $\IJcc=\IJaa$, and $C_J^{(2)}=C_J^{(1)}$, $I_c^{(2)}=I_c^{(1)}$,
see also Fig.~\ref{fig:IDSN}(a). 
Within certain ranges of the interface parameters, this structure supports an IDSN gate.
Similar to other ballistic gates that make use of the resonant forward-scattering,
it requires specific, large (shunt) capacitances $\gg C_J$ for the interface JJs. 
Unlike the NOT, ID, and NSWAP ballistic gates, the IDSN uses interface JJs 
which also have relatively large critical currents $> I_c$. 
For example, the particular IDSN gate of Ref.~\cite{OsbWus2018} 
has parameter values $(\CJaa,\IJaa) = (9 C_J, 2.4 I_c)$, 
$(\CJbb,\IJbb) = (21.3 C_J, 4.9 I_c)$, and 
$(\CJab,\IJab) = (6.0 C_J, 1.2 I_c)$,
while the geometric inductances in the interface are negligible, $\La=\Lb \ll L$.
Here we present a new IDSN gate with non-negligible interface inductance $\Lb$.
The parameters are given in the caption of Fig.~\ref{fig:IDSN}.
The new IDSN  differs in the details of the resonance dynamics from the one discussed in Ref.~\cite{OsbWus2018}; moreover, it has improved margins compared with the latter.

\begin{table}[tb]\centering
 \renewcommand\arraystretch{1.2}
\caption{
The logic table for the IDSN gate, whose allowed input states are null input (--), 
a single fluxon (0) or antifluxon (1) on either of the two input LJJs 
($S_1$ or $S_2$), 
or two synchronized fluxons of the same polarity coming in on both LJJs 
($S_1$ and $S_2$).  
The IDSN gate circuit is shown in Fig.~\ref{fig:IDSN}(a). 
An incoming fluxon on input LJJ $S_1$ ($S_2$) will be scattered to output LJJ 
$S_1'$ ($S_2'$), assuming the fluxon input velocity and synchronization lie 
in the acceptable range. 
}
\scalebox{0.9}{\sf
\begin{tabular}{|c|c|c|c|}
 \hline
 \multicolumn{2}{|c|}{Input} & \multicolumn{2}{c|}{Output} \\
 LJJ $S_1$  & LJJ $S_2$ & LJJ $S_1'$ & LJJ $S_2'$ \\
 \hline
\rowcolor[gray]{0.8} -- & -- & -- & -- \\
 0 & -- & 0 & -- \\
\rowcolor[gray]{0.8} -- & 0 & -- & 0 \\
 1 & -- & 1 & -- \\
\rowcolor[gray]{0.8} -- & 1 & -- & 1 \\
 0 & 0 & 1 & 1 \\
\rowcolor[gray]{0.8} 1 &  1 & 0 & 0 \\
 \hline
\end{tabular}
}
 \label{table:IDSN_logictable}
\end{table}

The fluxon dynamics for this new IDSN is shown in Fig.~\ref{fig:IDSN}(b-d). 
Subfigures (b) and (c) illustrate the dynamics for a single fluxon 
which is coming in on input LJJ $S_1$ and input LJJ $S_2$, respectively. 
The dynamics for these two initial conditions is of course equivalent, 
owing to the vertical symmetry of the structure. 
In either case, the incoming fluxon creates a short 
resonant excitation centered at the interface from which a new fluxon is 
created in the corresponding output LJJ, $S_1'$ and $S_2'$, respectively. 
The output fluxon has the same polarity as the input fluxon, 
similar to the 1-bit ID gate. 
However, the resonance here 
consists of a longer oscillation cycle at the interface than in 
the optimized 1-bit ID gate, cf.~Fig.~2 in Ref.~\cite{WusOsb2020}. 
The different dynamics in the IDSN can be attributed to the additional excitation 
of interface rail-JJs and evanescent field excitation in the 
non-fluxon carrying LJJs, see analysis in Sec.~\ref{sec:IDSN_equivalences}.

Subfigure (d) shows the scattering dynamics for two synchronized fluxons, 
each coming in on its own input LJJ. 
Here the dynamics in the upper and lower part of the structure is dynamically 
decoupled, because the current across the rail-JJ with phase $\phiBB$
cancels due to symmetry reasons. 
This is similar to the NSWAP gate with input fluxons of equal polarity, 
and as in that case the fluxons here also undergo NOT dynamics, 
i.e. they are forward-scattered to the output LJJs with inverted polarity.
Owing to the decoupled dynamics in upper and lower half, 
the resonance here is essentially identical to that of the 1-bit NOT gate, 
with however slightly different gate duration and efficiency. 
These differences are due to the deviation from the optimized parameters of the 1-bit NOT,
required here (as in 2-bit gates in general) 
for a compromise with the other IDSN operations.

\begin{figure*}[tb]
\includegraphics[width=\textwidth]{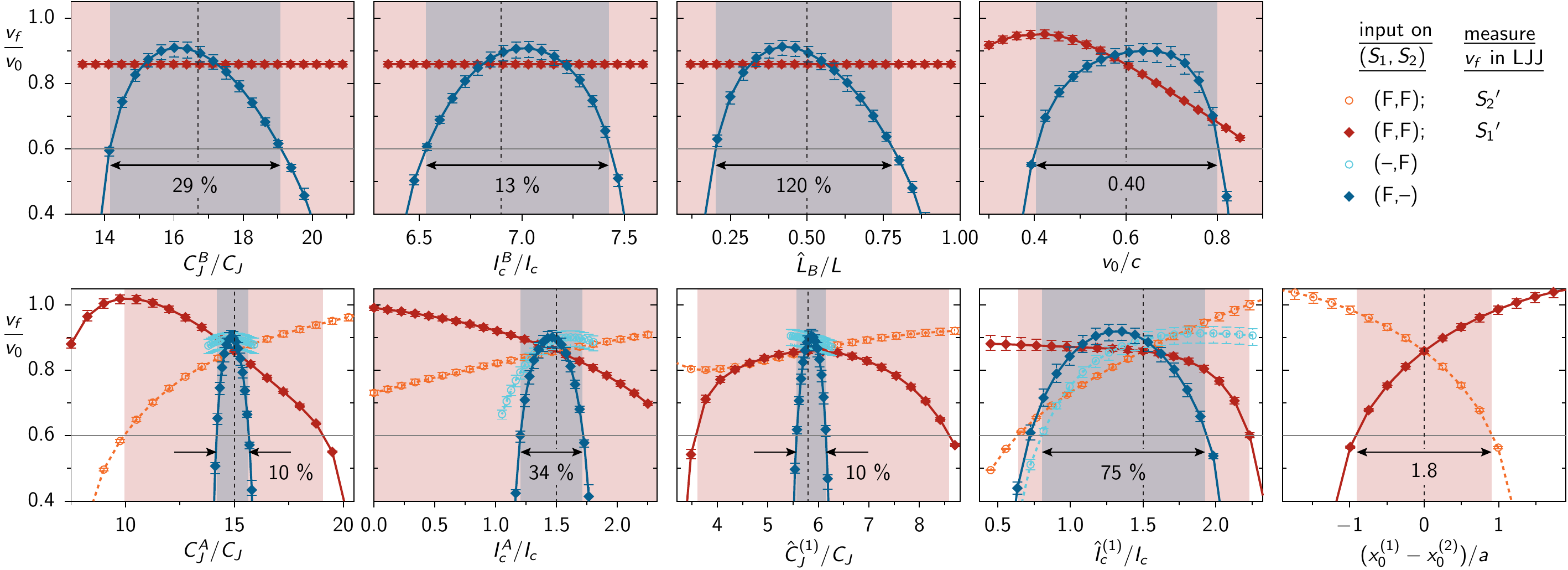}
\caption{
IDSN gate robustness and margins:
ratio of gate output to input velocity $v_f/v_0$ as function of interface 
parameters $\CJaa, \CJbb, \CJab^{(1)}, \IJaa, \IJbb, \IJab^{(1)}, \Lb$ 
(cf.~Fig.~\ref{fig:gateschematics}(b)), 
initial velocity $v_0 = v_0^{(1)} = v_0^{(2)}$, 
and fluxon separation $\Delta x = x_0^{(1)} - x_0^{(2)}$.
In each panel only a single parameter (interface parameter or 
initial state parameter $v_0$ or $\Delta x$) is varied, while all others are kept 
constant at the value indicated by the vertical dashed lines 
(cf.~caption of Fig.~\ref{fig:IDSN}).
The output velocity $v_f$ may depend on the input type:
(F,--) and (--,F) for single-fluxon input 
on LJJ $S_1$ ({\color{MidnightBlue} \large $\blackdiamond$})
and $S_2$ ({\color{SkyBlue} \large $\pmb{\circ}$}), respectively, 
and (F,F) for two-fluxon input, where $v_f$ refers to either the output fluxon 
on LJJ $S_1'$ ({\color{BrickRed} \large $\blackdiamond$})
or on LJJ $S_2'$ ( {\color{Orange} \large $\pmb{\circ}$}).
In the first row of panels, 
the variation of parameters $\CJbb, \IJbb, \Lb$ preserves the vertical symmetry 
of the structure,
as does the simultaneous variation of input velocities of both fluxons, $v_0$.
Thus, after the input of two fluxons (F,F) 
the output fluxons on both LJJs $S_1'$ and $S_2'$ 
have identical output velocity $v_f$.
Also, forward-scattering of a single fluxon on either of $S_1$ or $S_2$ 
leads to identical $v_f$.
In contrast, in the second row of panels the variation of interface parameters 
$\CJaa, \IJaa, \CJab^{(1)}, \IJab^{(1)}$ 
breaks vertical symmetry, 
and thus the final velocities $v_f$ of the two forward-scattered fluxons differ.
Similarly, a finite separation $\Delta x \neq 0$ of the two input fluxons 
in input type (F,F) breaks the symmetry of the initial state and 
thus $v_f$ of the two scattered fluxons differ.
Error bars indicate the amplitudes of velocity oscillations after scattering.
Shaded regions illustrate the ranges where $v_f/v_0 > 0.6$ is fulfilled 
for two-fluxon input (red) or single-fluxon input (blue), corresponding to 
fluxon energy conservation $E_{\text{fl},f}/E_{\text{fl},0} > 86 \%$.
Arrows indicate the effective margins resulting from this criterion. 
The admissible range for initial velocities is $0.4 c \leq v_0 \leq 0.8 c$, 
and the two fluxons in the (F,F)-type may be initially separated by up to $\pm 0.9$ cell, 
i.e.~the fluxons need to be synchronized with a delay time less than $0.09/\nu_J$.
}
\label{fig:IDSN_robustness} 
\end{figure*}

Fig.~\ref{fig:IDSN_robustness} summarizes the robustness characteristics of the IDSN gate
with respect to variations of the interface parameters and the initial state.
The individual panels of Fig.~\ref{fig:IDSN_robustness} show the ratio $v_f/v_0$ of output-to-input velocity of the forward-scattered fluxon(s) vs. the varied parameter, both for the single-fluxon processes (blue and light blue) and the two-fluxon processes (red and orange). 
Each panel shows the effect of one parameter variation around the optimized value (given in the caption of Fig.~\ref{fig:IDSN} and indicated here by vertical dashed lines), 
while all other parameters of the system are held fixed.
The parameter variations shown in the first row of panels preserve the top-bottom symmetry of the interface (variation of $\CJbb, \IJbb, \Lb$) or of the initial condition in the LJJs 
(equal variation of the fluxon input velocities in both LJJs, $v_0 = v_0^{(1)} = v_0^{(2)}$). 
Accordingly, the fluxon output velocity $v_f$ is here independent of whether 
the fluxon is sent in on input LJJ $S_1$ or $S_2$ in the single-fluxon process. 
Similarly, $v_f$ is equal in both output LJJs $S_1'$ and $ S_2'$ in the two-fluxon process.
That is why only two data lines are seen in the first row of panels.
In contrast, the parameter variations shown in the second row of panels 
break either the top-bottom symmetry of the interface 
(variation of $\CJaa \neq \CJcc, \IJaa \neq \IJcc, \CJab^{(1)} \neq \CJab^{(2)},  \IJab^{(1)} \neq \IJab^{(2)}$) 
or the top-bottom symmetry of the initial condition in the LJJs 
(variation of relative initial fluxon position $x_0^{(1)}-x_0^{(2)} \neq 0$ for double-fluxon process).
In these cases it therefore makes a difference whether a single fluxon 
is sent in on LJJ $S_1$ and measured in $S_1'$ (dark blue diamonds) 
or sent in on LJJ $S_2$ and measured in $S_2'$ (light blue circles). 
Similarly, in the two-fluxon process the output velocities measured on $S_1'$ (dark red diamonds) and $S_2'$ (orange circles) differ.

The shaded regions in Fig.~\ref{fig:IDSN_robustness} show the ranges where
$v_f/v_0 > 0.6$ is fulfilled, for either two-fluxon input (red shaded) or single-fluxon input (blue shaded), corresponding to an energy efficiency 
of $E_{\text{fl}}(v_f)/E_{\text{fl}}(v_0) > 86 \%$,
cf.~\Eq{eq:Efl_vdependence}.
The margins resulting from this efficiency criterion are indicated by the arrows.
Given current fabrication uncertainties, these margins are sufficiently wide 
to allow fabrication and testing.  
The interface parameters that need to be defined most precisely, within 10\%, 
are the capacitances $\CJaa,\CJcc$ of the upper and lower rail-JJs of the interface, 
and the capacitances  $\CJab^{(1)}, \CJab^{(2)}$ of the left and right interface JJs. 
In the two-fluxon process also the input fluxons need to be relatively well synchronized, 
with a delay time less than $0.09/\nu_J$ at velocity $v_0=0.6c$, 
corresponding to an admissible separation $x_0^{(1)}-x_0^{(2)}$ less than $0.9$ 
cells between the two fluxons.
The range of acceptable input velocities, $0.4 c \leq v_0 \leq 0.8 c$, is conveniently wide,
despite the resonant character of the underlying dynamics. 
This is different from the chaotic character of resonant scattering at point defects
in LJJs \cite{FeiKivVaz1992}.

\subsection{Equivalent 1-bit gate circuits}\label{sec:IDSN_equivalences}

We now briefly discuss how the IDSN gate operations are dynamically equivalent to 
certain 1-bit interfaces, depending on the input type.
The resulting mappings to approximately equivalent 1-bit interfaces
are summarized in table~\ref{table:equivalences__interfaceAA1BB1CC1_Npar13_20200130}
for the IDSN gate of Fig.~\ref{fig:IDSN}.
Similar reductions of the 2-bit gate dynamics have been discussed 
for the input cases of the NSWAP gate \cite{WusOsb2020}.

(i) 
When two fluxons of the same polarity approach the {\em vertically symmetric} 
2-bit interface at the same time, 
the currents excited in the rails of the interface form a vertically 
antisymmetric distribution, and the current through the center rail cancels. 
The dynamics in the upper and lower part of the interface is then effectively 
decoupled, and in each of them is equivalent to the dynamics of a 1-bit 
interface as shown in the left column
of table~\ref{table:equivalences__interfaceAA1BB1CC1_Npar13_20200130}.
The 1-bit interface has only a single rail-JJ,  
which is identical to one of the outer-rail JJs of the 2-bit interface, 
with parameters $(\CJaa,\IJaa)$. 

(ii)
Consider a single fluxon coming in on the upper input LJJ $S_1$. 
It dominantly excites the upper part of the IDSN structure,  
where the upper rail JJ starts winding by more than $\phiAA > \pi/2$.
Due to the current conservation on the interface rails, 
current also flows in the lower part of the interface.
The resulting phase fields in the lower LJJs, $S_2$ and $S_2'$,
have the form of exponentially localized edge states, 
and their amplitude remains small, $|\phi_n| \ll 1$. 
This allows us to approximately map the gate dynamics to that of a fluxon 
scattering in the 1-bit circuit shown in the right column of 
table~\ref{table:equivalences__interfaceAA1BB1CC1_Npar13_20200130}.
In this mapping each of the lower LJJs $S_2$, $S_2'$ of the IDSN, 
including its termination JJ with parameters $(\CJab, \IJab)$,
is replaced by a single effective JJ, with characteristics,
\begin{eqnarray}
\label{eq:Ceff_singlefluxonIDSN}
 C_J^{\alpha} &=& \CJab + \frac{C_J}{e^{2\mu a} - 1} \\
\label{eq:IJeff_singlefluxonIDSN}
 I_c^{\alpha} &=& \IJab + \frac{I_c}{e^{2\mu a} - 1} + \flqu \frac{(e^{\mu a} - 1)^2}{L (e^{2\mu a} - 1)}  \nonumber 
\,,
\end{eqnarray}
where $\mu$ is the inverse decay length ($\mu \lesssim 1/\lambda_J$) of the edge states. 
These quantities are derived in the Appendix, 
where we parametrize the LJJ fields by exponentially localized edge states, 
$\phi_n \propto e^{-\mu a |n|}$ and thus reduce the many degrees of freedom 
of each LJJ together with its termination JJ
to the amplitude of the edge state.
Furthermore, by comparing the plasma frequency of the effective JJ, 
$\omega_J^{\alpha}(\mu) = \sqrt{2\pi I_c^\alpha/(\Phi_0 C_J^\alpha)}$,
with the frequency $\omega_{\text{bulk}}(\mu)$ 
with which the edge states oscillate in the LJJ bulk, 
we can estimate $\mu = 0.68/\lambda_J$.
The resulting values from \Eq{eq:Ceff_singlefluxonIDSN}
are given in the right column of table \ref{table:equivalences__interfaceAA1BB1CC1_Npar13_20200130}.

We have simulated the fluxon scattering dynamics of the equivalent 1-bit interfaces
in table \ref{table:equivalences__interfaceAA1BB1CC1_Npar13_20200130}. 
In case (i) (left column of 
table \ref{table:equivalences__interfaceAA1BB1CC1_Npar13_20200130}), 
where the reduction is exact due to symmetry, 
there is of course full agreement with Fig.~\ref{fig:IDSN}(d). 
In case (ii) (right column of 
table \ref{table:equivalences__interfaceAA1BB1CC1_Npar13_20200130}),
although the reduction is only approximate, it nevertheless leads to excellent
agreement between the 2-bit gate dynamics, as is seen in the comparison 
between Fig.~\ref{fig:IDSN}(b) and Fig.~\ref{fig:IDSN_equiv_1bitgates}.

\begin{table}\centering
\renewcommand\arraystretch{2.0}
\caption{
Equivalent circuits of IDSN gate for the two fluxon input cases. 
For two synchronized input fluxons (left column) the current on the B-rail of the interface cancels,
such that the upper and lower part each are equivalent to a 1-bit (ID) gate.
For single fluxon coming e.g.~on upper input LJJ (right column), excitations in the lower left and right LJJs together with their parallel interface JJs  can be treated perturbatively, allowing to map to JJs $\alpha$.
IDSN parameters are those of Fig.~\ref{fig:IDSN}. 
The left and right termination JJs are identical in all interfaces,
$\CJab/C_J = 5.8$ and $\IJab/I_c = 1.5$.
}
\scalebox{0.9}{
\begin{tabular}{|c|c|}
\hline
  \parbox{3.7cm}{\centering \bf Two synchronized fluxons (same polarity)}
& {\bf Single fluxon} \\
  \includegraphics[width=3.0cm]{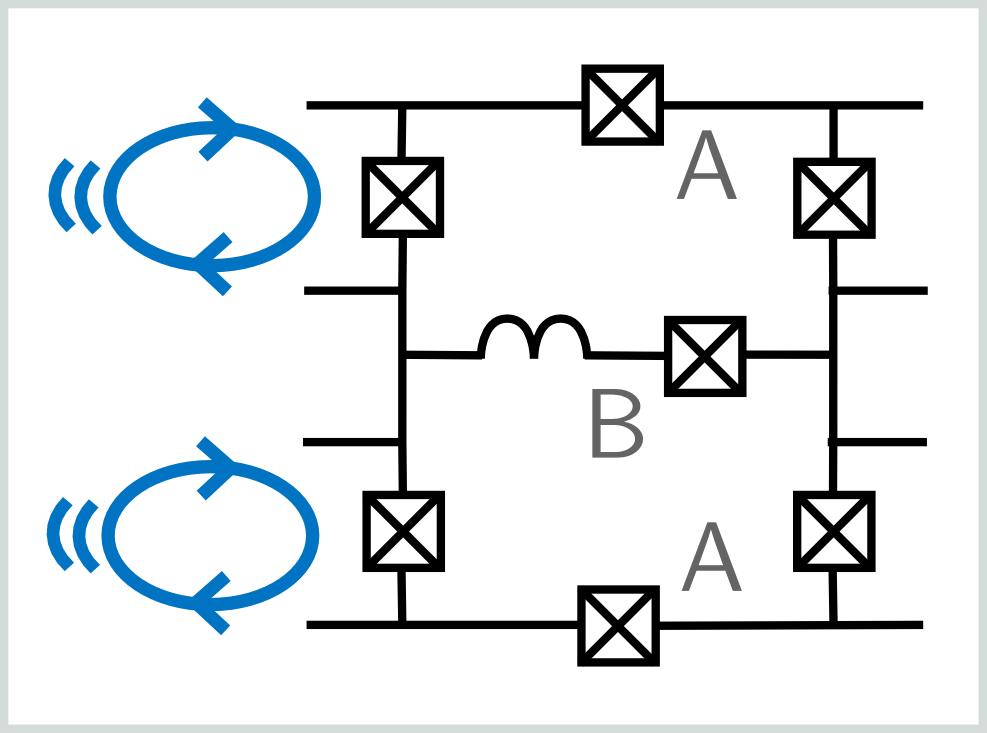}
& \includegraphics[width=3.0cm]{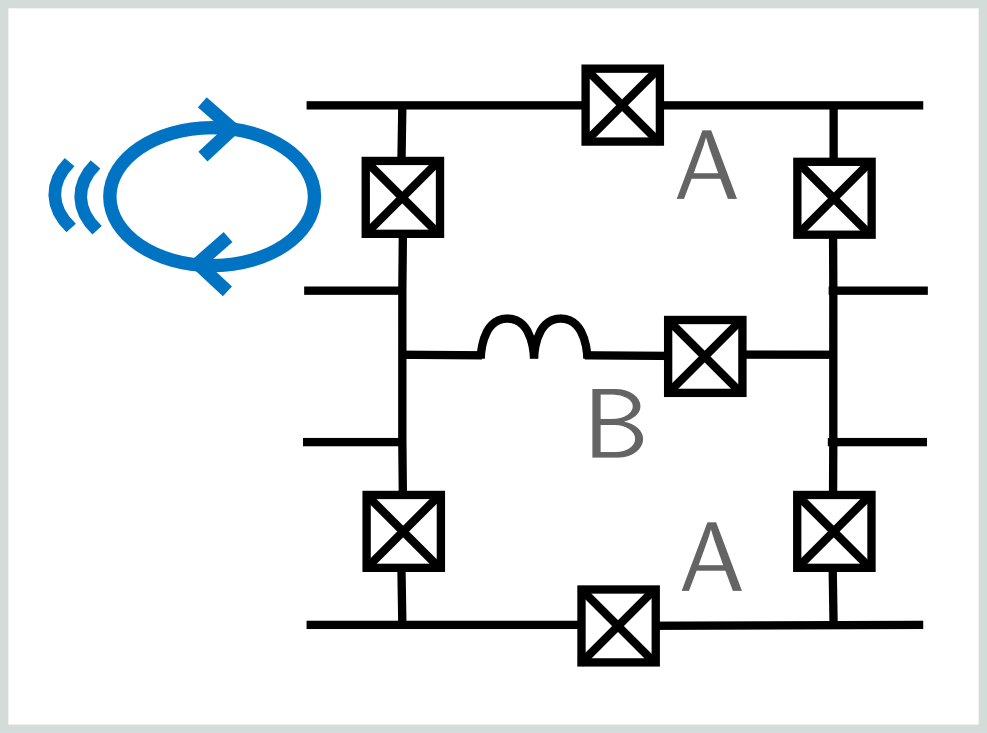} \\
\multicolumn{2}{|c|}{
 \parbox{6.5cm}{
e.g. with parameters\newline
 A,C-rail: $\CJaa/C_J = 15.0$, $\IJaa/I_c=1.5$ \newline 
 B-rail: $\CJbb/C_J = 16.7$, $\IJbb/I_c=6.9$, $\Lb/L = 0.5$\newline 
 }} \\ 
\multicolumn{2}{|c|}{{\bf Equivalent 1-bit interfaces:}} \\
  \includegraphics[width=3.0cm]{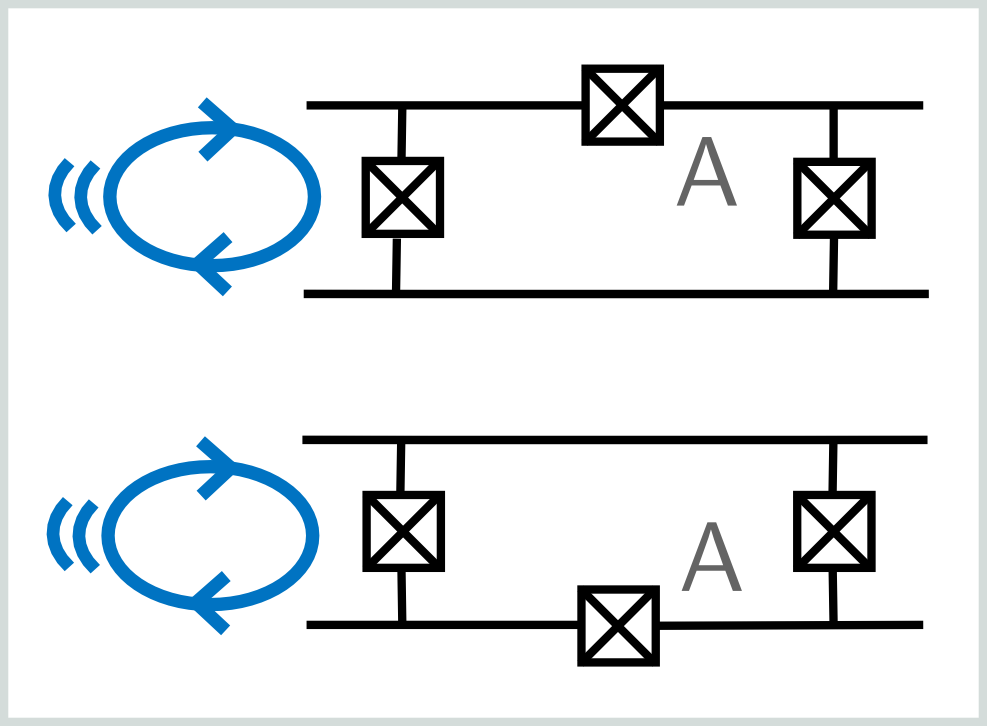}
& \includegraphics[width=3.0cm]{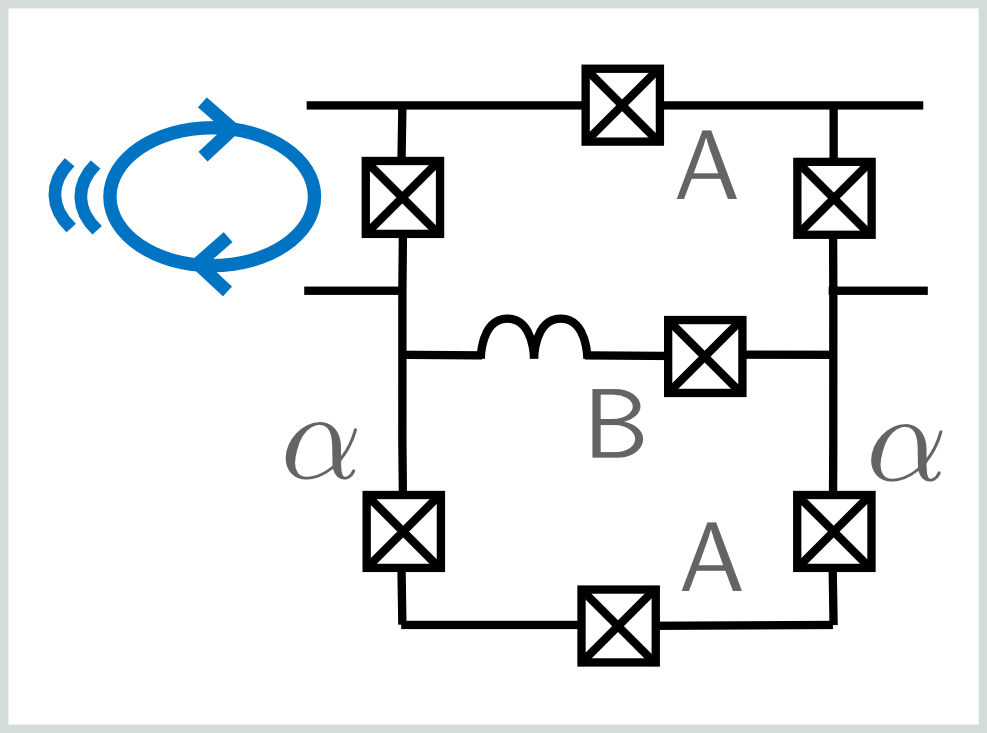} \\
\parbox[t]{3.7cm}{
$\CJaa/C_J = 15.0$, $\IJaa/I_c = 1.5$\\
}
&
\parbox[t]{3.7cm}{
$\CJaa/C_J = 15.0$, $\IJaa/I_c = 1.5$\\[1ex]
$\CJbb/C_J = 16.7$, $\IJbb/I_c = 6.9$\\[1ex]
$C_J^{\alpha}/C_J \approx 7.3$, $I_c^{\alpha}/I_c \approx 3.9$\\
} \\
 \hline
\end{tabular}
}
\label{table:equivalences__interfaceAA1BB1CC1_Npar13_20200130}
\end{table}

\section{A store and launch (SNL) gate}\label{sec:SNL}

\begin{figure}[tb]\centering
 \includegraphics[width=\columnwidth]{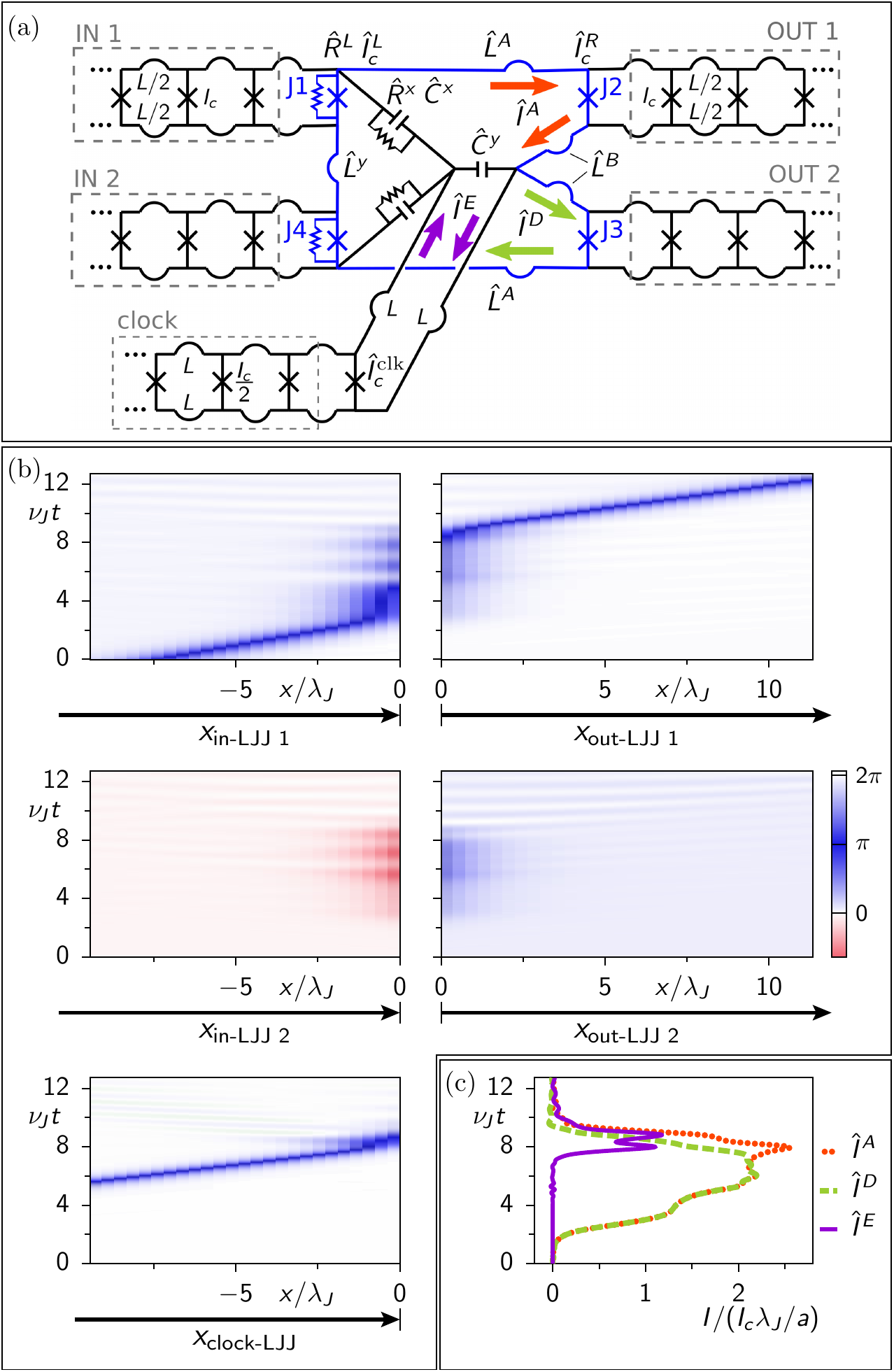}
 \caption{
  The 2-input Store-and-Launch (SNL) gate 
  receives one input data fluxon from one of two possible input LJJs. 
 (a) SNL gate circuit. 
  After the data fluxon enters through input LJJ 1 or 2, the bit is stored
  as a circulating current in a center storage cell which contains JJs J1-J4 
  (connected to the input and output LJJs)
  and inductors $\La, \Lb$, etc..
  The storage cell (marked in blue) has bottom-top symmetry, where the symmetry 
  line is defined by the center capacitor $\hat{C}^{y}$ and connected clock LJJ.
 (For clarity, some circuit elements are left unlabeled, 
 such as $\hat{I}_c^{L}[\text{J4}]=\hat{I}_c^{L}[\text{J1}]$, 
 because they are determined by top-bottom symmetry.)
  A low-energy clock fluxon is sent in on the clock LJJ
  and will cause a stored bit state 0 (1) to be launched as a fluxon (an antifluxon)  
  on the upper (lower) output LJJ.
 (b) LJJ phases $\phi_n(t)$ vs positions $x$ and time $t$:
 A data fluxon (bit state 0) comes in on input LJJ 1 
 with velocity $v_{\text{in}}=0.4c$
 and settles into the storage cell of the SNL. 
 The stored data bit later gets launched into output LJJ 1 by a low-energy clock fluxon 
 arriving from the clock LJJ at $\nu_J t_{\myclock} \approx 8$. 
 (c) Storage-cell currents $\hat{I}(t)$, as defined in (a):
 The current generated by the stored data fluxon is equal in upper and lower parts of the storage cell, $\hat{I}^D \approx \hat{I}^A$ for $t<t_{\myclock}$.
 Later, the superposition of this storage current with that of the clock fluxon 
 (with same polarity as the data fluxon, $\hat{I}^E \cdot \hat{I}^A > 0$)
 leads to a current imbalance $\hat{I}^A > \hat{I}^D$,
 and eventually to the launch of the stored bit as a new fluxon into output LJJ 1. 
 In contrast, a stored antifluxon (bit state 1) would get launched into output LJJ 2.
 The parameters of the simulation for (b,c) are given in the text.
 }
 \label{fig:SNL}
\end{figure}

The IDSN gate is an example of a ballistic 2-bit gate, with an energy efficiency close to unity 
($ \geq 86 \%$ for the IDSN gate with margins defined 
in Fig.~\ref{fig:IDSN_robustness}).
In the case of its two-fluxon operation, the high energy efficiency however relies on the input of two well-synchronized input fluxons, cf.~last panel in Fig.~\ref{fig:IDSN_robustness}. 
Moreover, fluxon velocity of course needs to be restored after a sequence of ballistic gates. 
Therefore, in addition to the {\em ballistic logic} gates, 
{\em clocking} gates are required for fluxon synchronization 
and for restoring fluxons to a velocity within the velocity range of the ballistic gates. 
We have developed such Store-and-Launch (SNL) gates,
which store the bit state of an incoming (slowed-down) {\em data} fluxon. 
Later, triggered by the interaction with a timed {\em clock} fluxon,
the SNL launches the stored bit as a {\em data} fluxon carrying the original bit state
on an output LJJ (at a certain higher speed). 
The clock fluxon, which is annihilated in the process, serves as the power source
of this thermodynamically irreversible gate.

Earlier we had introduced a 1-input SNL gate, which has a single input LJJ for data \cite{OsbWus2018}. 
Here we report on a related 2-input SNL, where a data fluxon can come in 
on one of two input LJJs. 
The schematic of the 2-input SNL is shown in Fig.~\ref{fig:SNL}(a). 
As an additional feature, our SNL gates have two output LJJs.
The output fluxon is launched into one of them, depending on the stored bit state. 
Both, the bit-dependent routing and the existence of more than one input port, 
are not general properties of clocking gates, but are required 
in the particular application for which we have developed the SNL gate,
namely as a component of a CNOT gate. 
The application is discussed below in Sec.~\ref{sec:CNOT}, including Fig.~\ref{fig:CNOT_schematics}.

%

The gate circuits of the {\em ballistic} RFL gates
have left-right symmetry as a necessary condition for logic reversibility:
running the gate `backward', by reverting the momentum of the output fluxons, will restore the gate's input state. 
In a ballistic gate the sets of input and output LJJs are therefore interchangeable.
In contrast, the irreversible SNL gates do not have left-right symmetry, 
and the roles of each LJJ is fixed as either input or output channel.
The SNL gates, however, have top-bottom symmetry, such as in the 2-input SNL of Fig.~\ref{fig:SNL}(a),
where the center capacitor $\hat{C}^y$ together with the clock LJJ (coming out of the paper) define the symmetry plane of the circuit.

The 2-input SNL circuit, Fig.~\ref{fig:SNL}(a), consists of a central storage cell, 
comprised of inductors of small values $\La,\Lb,\hat{L}^y$, and the JJs J1--J4, 
with critical currents $\hat{I}_c^L$, $\hat{I}_c^R$, which terminate the input and output LJJs.
On the input side these JJs (J1 and J4) are also resistively shunted with resistance $\hat{R}^L$. 
All other JJs in Fig.~\ref{fig:SNL}(a) are undamped, and all JJs have the same plasma frequency as the JJs in the LJJs, $\omega_J = \sqrt{2\pi I_c/(\Phi_0 C_J)}$.
Additional circuit elements are either directly in parallel to the clock LJJ, such as capacitance 
$\hat{C}^y$, or in parallel with other parts of the storage cell, such as capacitance $\hat{C}^x$ and resistance $\hat{R}^x$.

All input and output LJJs have equal properties, with the JJ characteristics $(C_J, I_c)$, 
cell inductance $L$, and cell length $a$.
The parameters of the clock LJJ are scaled relative to these data LJJs according to
$(C_J^{\myclock}, I_c^{\myclock}, L^{\myclock}, a^{\myclock}) = (C_J/s, I_c/s, s L, a)$,
here with a factor $s=2$.
This scaling implies that the clock LJJ has the same Josephson penetration depth $\lambda_J$, and the same upper velocity bound $c=\omega_J \lambda_J$ as the data LJJs. 
However, compared with the latter ones, 
the characteristic impedance in the clock LJJ is a factor of $s$ larger, 
$\sqrt{L^{\myclock}/C_J^{\myclock}} = s \sqrt{L/C_J}$,
and the rest energy of a clock fluxon is reduced by a factor of $s^{-1}$,
$\Efl^{\myclock}(0) = \Efl(0)/s$.
As will be discussed in more detail later, this is done in order to minimize the energy cost of the launching process, in which the clock fluxon is annihilated.
Its energy should therefore be as small as possible (while still properly launching the data fluxon) to make the SNL energy-efficient.

Fig.~\ref{fig:SNL} illustrates the dynamics of the 2-input SNL in simulation. 
The panels of subfigure (b) show colormaps of the JJ-phases $\phi_n(t)$ vs position $x$ 
in the individual LJJs: 
$x=0$ corresponds to the position of the storage cell, 
$x < 0$ is assigned to the input and clock LJJs to the left,
and $x > 0$ to the output LJJs to the right. 
Subfigure (c) shows the currents in the storage cell 
and at the end of the clock LJJ.
When a data fluxon arrives from one of the input LJJ(s) 
-- here it is the upper one -- 
it generates a current circulating in
the storage cell,
with evanescent excitations of the JJs in the input and output LJJs. 
The current distribution over the various inductors differs 
from that of a soliton in the LJJ, 
but of course the total flux is still one flux quantum. 
The circulation direction of the storage current is equal to that of the
incoming data fluxon: for input bit state 0 (1) it circulates clockwise (anticlockwise).
Due to the geometry of the SNL, 
the storage current is distributed symmetrically in the storage loop, 
$\hat{I}^A \approx \hat{I}^D$
(indicated by red and green arrows in Fig.~\ref{fig:SNL}(a) 
and corresponding data line in subfigure (c)).
The symmetry implies that no current $\hat{I}^E$ (ac or dc) 
is excited at the end of the clock LJJ 
(purple arrows and data line) during this storage stage of the operation. 

Later, a clock fluxon is sent in through the clock LJJ and arrives at time $t_{\myclock} \approx 8/\nu_J$. 
The clock fluxon, which is assumed to be of polarity equal to data state 0,
induces a current $\hat{I}^E$ (purple arrows) at the storage cell, 
which adds constructively (destructively) with the current $\hat{I}^A$ ($\hat{I}^D$) 
in the upper (lower) part of the storage cell (red and green).
This results in an imbalance, $\hat{I}^A > \hat{I}^D$, which grows 
and feeds the phase gradient in the upper output LJJ 1. 
The phase gradient in the upper output LJJ 1 gradually forms into a new data fluxon which eventually launches.
If the storage current were circulating anticlockwise instead, 
corresponding to a stored bit state 1, 
an antifluxon would get launched, but then into the lower output LJJ 2. 
After the launching process, the currents in the storage cell 
and in the clock LJJ go to zero since flux is no longer stored.

When the SNL gate stores flux, it stores a large fraction of the rest
energy (potential energy) of the incident fluxon. 
However, some energy of the input fluxon is dissipated in the resistors
to ensure that the fluxon 
is not immediately emitted from the storage loop back to one of the input LJJs, 
but settles into a static stored flux state.
In the simulation shown in Fig.~\ref{fig:SNL}
the damping occurs as a two-step process, with temporary dissipation events 
at $\nu_J t \approx 3$ (when the incoming data fluxon is initially stopped close to $x \lesssim 0$) 
and at $\nu_J t \approx 6$ (when the data fluxon is finally absorbed fully such that the storage current rises to a saturation value). 
Later, when the clock fluxon arrives, it generates dissipative currents mostly on the bridge resistors $\hat{R}^x$. Dissipation is dominant on the lower of these two 
in the example shown in Fig.~\ref{fig:SNL} 
where the storage current circulates clockwise (bit state 0) 
and a fluxon is launched into output LJJ 1.

One purpose of our SNL clocking gate is to restore the energy of a data fluxon 
to a value suitable for the input fluxon of a ballistic gate like the IDSN.
For the SNL gate studied in Fig.~\ref{fig:SNL} we use a data fluxon with input velocity $v_{\text{in}} = 0.4 c$, 
corresponding to $\Efl(v_{\text{in}}) = 8.7 E_0$, cf.~\Eq{eq:Efl_vdependence}.
This value is close to the minimum of the operational range of the SNL.
(An input fluxon with velocity below this minimum would get reflected instead of being stored.)
The bit gets stored as a flux quantum in the storage cell 
with an energy of $U_{\text{stored}} \approx 8 E_0$, 
close to the rest energy of the data fluxon. 
The clock fluxon is sent in 
with $v^{\myclock} = 0.6c$ in the simulation of Fig.~\ref{fig:SNL},
corresponding to the fluxon energy $\Efl^{\myclock}(v^{\myclock}) = 5 E_0$ 
(recall that the rest energy of a clock fluxon is smaller than that of a data fluxon,
$\Efl^{\myclock}(0) = \Efl(0)/2 = 4 E_0$). 
The initial energy of the input fluxons is 
$E_{\text{init}} = \Efl(v_{\text{in}}) + \Efl^{\myclock}(v^{\myclock}) = 13.7 E_0$. 
In the simulations the output data fluxon has an energy of 
$\Efl(v_{\text{out}}=0.51 c) = 9.3 E_0$, 
giving an energy efficiency of $\Efl(v_{\text{out}})/E_{\text{init}} = 69 \%$. 
Even though this is less efficient compared with the ballistic RFL gates, 
it is efficient relative to irreversible logic (CMOS or irreversible SFQ), 
which consumes at least the entire potential energy of the bit state in switching.

The SNL energies and efficiencies are summarized in the 2nd column of table \ref{table:SNL_efficiencies},
both for the case of  $v_{\text{in}}=0.4c$ (shown in Fig.~\ref{fig:SNL})
and the case of a larger input velocity, $v_{\text{in}}=v_{\text{out}}$,
where the efficiency drops to $65 \%$.
One may compare them to the values that are theoretically achievable in an ideal case.
We define an `ideal' SNL as one where the energy of the new data fluxon 
equals the sum of the original data fluxon's rest energy and the energy of the clock fluxon,  
$\Efl(v_{\text{out}}) = 8 E_0 + \Efl^{\myclock}(v^{\myclock})$.
In other words, the clock fluxon transfers its entire energy to launch a fluxon from the stored bit state, which is stored at an energy $U_{\text{stored}} = 8 E_0$ corresponding to that of a static fluxon in the LJJ bulk.
This choice of $U_{\text{stored}}$ ensures that input data fluxons of (in principle) arbitrary input velocity $v_{\text{in}}$ could be successfully stored. 
(Whereas in the realistic case of Fig.~\ref{fig:SNL} a lower $v_{\text{in}}$-bound exists, as mentioned earlier.)
The 3rd and last column of table \ref{table:SNL_efficiencies} list the resulting maximum efficiencies, 
for $v^{\myclock}=0.6 c$
and two clock LJJ scaling factors $s = \Efl(0)/\Efl^{\myclock}(0)$.
The ideal energy efficiency (at $v_{\text{in}} = v_{\text{out}}$) 
for the SNL gate with $s=2$ is $72\%$ (3rd column).
The 1-input SNL introduced earlier \cite{OsbWus2018} uses $s=4$, 
and here we find the ideal efficiency 
(again at $v_{\text{in}} = v_{\text{out}}$) for this gate to be $81\%$ (last column).

\begin{table}[tb]\centering
 \caption{
 SNL gate performance, which depends on $s$ (clock LJJ scaling), 
 $v^{\myclock}$, 
 and $v_{\text{in}}$. 
 Relations $\Efl(v)$ are according to \Eq{eq:Efl_vdependence}, and 
 the efficiency is 
 $\Efl(v_{\text{out}})/(\Efl(v_{\text{in}}) + \Efl^{\myclock}(v^{\myclock}))$.
`Ideal' performance assumes that the bit state is stored at the rest energy 
$8 E_0$ of the data fluxon, i.e. its kinetic energy is lost, 
and that the entire clock fluxon energy is transferred to the launched fluxon
(see text).
 }
 \renewcommand{\arraystretch}{1.2}
 \scalebox{1.0}{
 \begin{tabular}{|p{2.1cm}|p{1.6cm}|p{1.6cm}|p{1.6cm}|}
  \hline
  & parameters & \multicolumn{2}{l|}{`ideal': } \\
 & of Fig.~\ref{fig:SNL} & \multicolumn{2}{c|}{
 $\Efl(v_{\text{out}}) = 8 E_0 + \Efl^{\myclock}(v^{\myclock})$}\\
  \hline
  \rowcolor[gray]{0.8} s & 2 & 2 & 4 \\
  \rowcolor[gray]{0.8} $\Efl^{\myclock}(v^{\myclock}=0.6c)$ & $5 E_0$ & $5 E_0$ & $2.5 E_0$ \\
  $\Efl(v_{\text{out}})$ & $9.3 E_0$ & $13 E_0$ \hspace*{0.1cm}(**) & $10.5 E_0$ \hspace*{0.1cm}(**)\\
  $v_{\text{out}}$ & $0.51 c$ \hspace*{0.3cm}(*) & $0.79 c$ & $0.65 c$  \\
  \hline
  \hline
  \rowcolor[gray]{0.8}
  \multicolumn{4}{|l|}{Slow input case: $v_{\text{in}} = 0.4c$ } \\
  $\Efl(v_{\text{in}})$ & $8.7 E_0$ & $8.7 E_0$ & $8.7 E_0$ \\
  efficiency & 69\% & 95\% & 94\% \\
  \hline
  \rowcolor[gray]{0.8}
  \multicolumn{4}{|l|}{Fast input case: $v_{\text{in}} = v_{\text{out}}$ } \\
  $v_{\text{in}}$ & $0.51 c$ & $0.79 c$ & $0.65 c$ \\
  $\Efl(v_{\text{in}})$ & $9.3 E_0$ & $13 E_0$ & $10.5 E_0$ \\
  efficiency & 65\% & 72\% & 81\% \\
  \hline
  \end{tabular}  
  }\\[0.5ex]
  (*) measured in the simulation \hspace*{0.5cm} (**) according to stated criterion
 \label{table:SNL_efficiencies}
\end{table}

For the 2-input SNL circuit of Fig.~\ref{fig:SNL}(a)
we have tuned the parameters of the storage cell
for an efficient conversion of the clock-fluxon energy into kinetic energy of the launched fluxon.
The optimization is done for fixed clock fluxon energy and 
under the condition that the input energy of the data fluxon lies in a range 
that covers at least the interval $(8.7 E_0, 10 E_0)$.
Using these criteria we find the following parameters of the storage cell:
$\hat{I}_c^{L}/I_c = \hat{C}_J^{L}/C_J = 2.6$, $1/\hat{R}^{L} = 0.1 / Z$
(JJs J1 and J4),
$\hat{I}_c^{R}/I_c = \hat{C}_J^{R}/C_J = 2.4$ (JJs J2 and J3),
$\La=\Lb=\hat{L}^{y} \ll L$, 
$\hat{C}^x = 2.0 C_J$, $1/\hat{R}^x = 1.7 / Z$, $\hat{C}^y = 1.0 C_J$ 
(storage cell with damping elements and coupling to clock LJJ).
Herein $Z=\sqrt{L/C_J}$ is the characteristic impedance of the data LJJs.
As mentioned above, the clock LJJ has bulk parameters  
$I_c^{\myclock}/I_c =  C_J^{\myclock}/C_J = 0.5$ and
$L^{\myclock}/L = 2.0$;
however the last JJ is here also modified 
$\hat{I}_{c}^{\myclock}/I_c = \hat{C}_{J}^{\myclock}/C_J = 1.0$.  
These parameters of the 2-input SNL gate are used in the simulation shown in Fig.~\ref{fig:SNL}.

We note that we have found parameters with improved energy efficiency compared with the parameters used in Fig.~\ref{fig:SNL}.
However, these usually only allow a very narrow range of input velocities: 
If the input velocity lies below that range, the data fluxon gets reflected 
before entering the storage cell. 
If the velocity exceeds the upper limit of that range, the input data fluxon 
is also not stored but moves directly into the other input LJJ. 
In contrast, the previously studied 1-input SNL \cite{OsbWus2018}
does of course not have the latter limitation.
Besides, the 1-input SNL can have a higher launch efficiency
because the stored current excites evanescent phase fields in one fewer LJJ.
In some applications, however, it is necessary to have an SNL gate 
with the same number of input and output LJJs (as well as bit dependent routing).
One important example is the CNOT gate discussed in the following section.

\section{A CNOT gate in RFL}\label{sec:CNOT}

\begin{figure}[tb]\centering
\includegraphics[width=0.8\columnwidth]{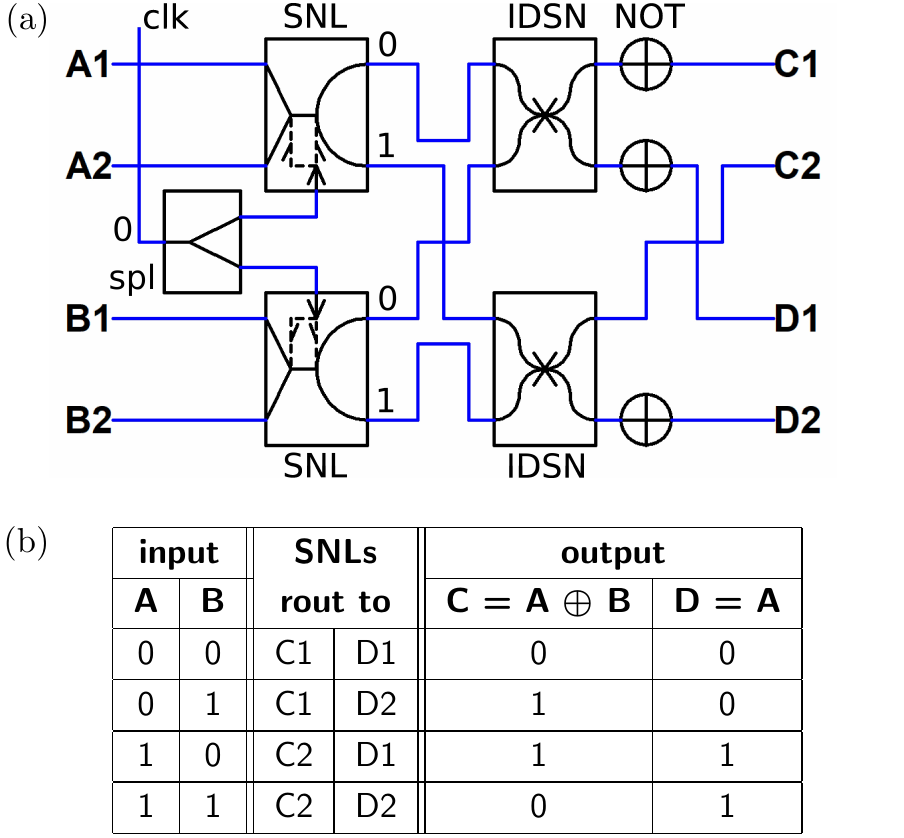}
\caption{
(a) Schematic for a composite CNOT logic gate designed with RFL components:
two ballistic IDSN gates, two 2-input SNL clocking gates, three NOT gates, 
a splitter, and LJJs as ``wiring'' (blue).
(b) CNOT logic table resulting from gate actions and routing in (a).
The bit states A and B (columns 1,2) are defined by the two data fluxons 
coming in on one of the input ports A1 or A2, and on one of B1 or B2, respectively.
The input bit states A and B are each stored in an SNL gate.
A clock fluxon is sent in and at a splitter fans out into two half-energy ones.
These clock fluxons power the synchronized launch of the stored data bits. 
The launched data bits are routed on bit-state dependent paths (columns 3,4).
After action of the IDSN and NOT gates 
the bit state D (appearing either on path D1 or D2) is a copy of A, 
and the bit state C (on path C1 or C2) is XOR(A,B).
}
 \label{fig:CNOT_schematics}
\end{figure}

Some of our 2-bit logic gates are purely ballistic, i.e. the logic action
is generated alone through the inertia of (synchronous) input fluxons.
Examples include the NSWAP \cite{WusOsb2020} and IDSN gate. 
However, fluxon dynamics in the 2-bit gate circuit, Fig.~\ref{fig:gateschematics}(b),
can only render a subset of all reversible 2-bit logic gates:
namely those which act in the same way on an input state combination as on its inverse, e.g. on the input state (0,0) as on (1,1).
Clearly, the NSWAP gate and IDSN gate (see table~\ref{table:IDSN_logictable}) belong to this subset. 
The above restriction originates from the absence of an external magnetic field 
or stored flux, making the circuit dynamics invariant 
under a sign change of the phases.
Additionally, 
a sign change of the input LJJ phases corresponds to the inversion of the input 
fluxon polarities, and thus to the inversion of the input bit states.
Thus the gate dynamics resulting e.g. for the input state (0,0) is fully equivalent 
to that for input state (1,1). 

Other reversible logic gates, such as the CNOT do not belong to 
the above-mentioned subset and thus cannot be achieved by a ballistic RFL gate alone.
However, by combining ballistic logic gates with circuit elements for synchronization and routing we can construct more complex logic gates such as the CNOT \cite{OsbWus2018}. 
Fig.~\ref{fig:CNOT_schematics}(a) shows the schematic for an RFL implementation of a CNOT, which consists of two SNL clocking gates, two IDSN logic gates, 
and other routing components.

We shortly describe the operation of the composite CNOT:
The input consists of two fluxons, carrying bit states A and B, respectively.
Bit A comes in from the left on either of the input ports A1 or A2
of a 2-input SNL where it is stored.
Similarly, bit B comes in on either B1 or B2 and is stored in a second SNL. 
Later a clock fluxon is sent in on a clock LJJ, 
which has regular LJJ characteristics $(C_J, I_c, L, a)$.
A T-branch splitter \cite{phasemodelogic1, phasemodelogic2} divides the original clock LJJ into two clock LJJs with characteristics $(C_J/2, I_c/2, 2 L, a)$, and as a result the original clock fluxon splits into two half-energy clock fluxons moving at the original speed. 
The two clock LJJs are each connected to a SNL, 
where the impinging clock fluxons synchronously launch the stored bit A and B as a fluxon or antifluxon. 
The launch is bit-state dependent, where the bit state 0 (1) is launched 
as a fluxon (antifluxon) into the upper (lower) output LJJ of each of the SNLs. 
We note that the two clock LJJs have to be wired to their respective SNL in a different way, 
in order to achieve the same launch directionality in both LJJs. 
This difference is indicated in Fig.~\ref{fig:CNOT_schematics} by the two distinct circuit symbols for the two SNLs. 
While the upper SNL corresponds exactly to the schematic of Fig.~\ref{fig:SNL}(a), the lower SNL 
has a half twist in the clock LJJ relative to that schematic.

The output LJJs of the two SNLs are connected to two IDSN gates in such a way 
that any bit state 0 of the two launched data fluxons 
will arrive at the upper IDSN and any bit state 1 will arrive at the lower IDSN.
Depending on the initial bit states (A,B), 
either both IDSN gates receive a single fluxon as their input,
or only one IDSN gate receives two input fluxons with the same polarity.
According to the logic action of the IDSN, table~\ref{table:IDSN_logictable}, 
in the former case the single input fluxon or antifluxon undergoes an ID operation. 
In the latter case, the two fluxons are well synchronized thanks to the simultaneous launch from the two SNLs. 
They will then both undergo NOT dynamics in the IDSN gate. 
The output fluxon(s) of the IDSN appear on its upper (lower) output LJJ 
if they entered on the upper (lower) input LJJ.
Next, some of the fluxons pass through a NOT gate on the way to the output ports C1, C2, D1, D2 of the CNOT. 
The routing to these ports is done such that 
exactly one fluxon arrives on either C1 or C2, 
and thus uniquely defines the output bit C of the CNOT gate. 
Similarly, 
exactly one fluxon arrives on either D1 or D2 and defines the output bit D. 
The combined action of the clocking gates (SNL), ballistic logic gates (IDSN, NOT), and the routing ensures that these output bit states are related to the input bit states in form of the CNOT logic, 
see table in Fig.~\ref{fig:CNOT_schematics}(b).
These bits can then each be stored in another SNL for subsequent processing if desired.
For example, one can cascade two CNOTs after one another to logically reverse the operation.

\section{Conclusion}

Reversible digital logic is important for the future development of computing 
because of the much higher energy efficiency compared with irreversible logic gates,
including industry CMOS gates and the most developed SFQ logic. 
Recent demonstrations of such gates in superconducting circuits
are {\em adiabatic}-reversible, 
making use of the inverse scaling between gate time and energy cost.
In contrast, we are developing Reversible Fluxon Logic (RFL) based on 
{\em ballistic}-reversible gates. 
These are powered alone by the inertia of incoming fluxons which at the same time  
are the input bits themselves.
The ballistic gates in RFL exploit a resonant scattering process at special interfaces between LJJs.
This scattering achieves conditional polarity inversion of an input fluxon, 
dependent on the gate type and input bits.  
The advantage of RFL over irreversible logic is the full conservation of the rest (potential) energy
of fluxons in logic operations. 
In our simulations this makes up $80\%$ of the fluxons' total energy. 
Ballistic-reversible gates moreover conserve a large fraction of the kinetic fluxon energy, thus yielding conservation of up to $97\%$ of fluxon energy.
Ballistic-reversible gates include the 1-bit NOT and ID, and the 2-bit NSWAP and IDSN gates.

The CNOT is implemented in RFL as a composite gate from a set of ballistic and non-ballistic gates.
This includes two IDSN gates, three fundamental NOT gates, and two Store-and-Launch (SNL) gates. 
The SNL is a clocking gate that stores the bit state of an incoming fluxon
and later launches it as a new fluxon on a bit-state dependent output LJJ.
In the SNL version described here, 
the launch is powered by a clock fluxon at half of the data fluxon energy.
In the CNOT, the data fluxon launch from the two SNL gates is synchronized, 
using the simultaneous arrival of clock fluxons which are fanned out from a single source.

Unlike the ballistic RFL gates which are nominally undamped the SNL gate uses damping resistors
to ensure that the incoming data fluxon gets stored 
as a static flux in the SNL storage cell.
In the launch process the clock fluxon is annihilated.
A part of its energy goes to the launched data fluxon, which then may end up 
with larger energy than the input data fluxon.
The other, larger part of the clock fluxon energy is dissipated in the resistors. 
However, since the clock fluxon has only a fraction of the data fluxon's energy,
and since much of the rest energy of the input data fluxon is preserved,
the SNL gate may be very efficient by irreversible logic standards.

These studies indicate that reversible computing should not be 
thought of only in terms of the adiabatic model. 
Ballistic-reversible gates enabled by resonant fluxon scattering 
seem to be an implementable alternative.
Combined with clocking gates that preserve the rest energy of the bits,
ballistic gates can be used to build complex logic gates.

\section*{Acknowledgements}
KDO would like to thank Quentin Herr, Mike Frank, Naoki
Takeuchi, Nobuyuki Yoshikawa, Gary Delp and Ivan Sutherland for 
discussions on logic gates. 
KDO and WW would also like to acknowledge Liuqi Yu at LPS 
for scientific discussions on experimental designs of RFL.
WW would like to thank the Physics department 
at the University of Otago for its hospitality.

\appendix[Modelling an LJJ with weakly excited edge state]
\label{sec:boundstatemodel_singleLJJ}

In the single-fluxon operation of the IDSN gate, where e.g. the fluxon 
comes in and leaves on LJJs $S_1$ and $S_1'$, the two other LJJs $S_2$ and $S_2'$ 
are only weakly excited throughout the gate operation, cf.~Fig.~\ref{fig:IDSN}(b). 
The phase fields temporarily excited in these LJJs have the form of exponentially 
localized edge states at the interface and undergo a slow coherent oscillation.  
The contribution of each of those LJJs -- including the parallel interface JJ -- 
can then be analyzed in terms of a model,
which parametrizes the phase fields as edge states, 
e.g. on the lower right interface JJ and LJJ $S_2'$,
\begin{equation}\label{eq:boundstate_rightLJJ}
 \phi_n = \phi_R(t) e^{-\mu a n} \,,
\end{equation}
with an inverse decay length $\mu$. 
Herein, $n=0$ labels the lower right interface JJ with characteristics $(\CJab,\IJab)$, 
and $n=1,2,\ldots$ label the JJs in $S_2'$ in increasing distance 
$x = a n$ from the interface.

\begin{figure}[tb]\centering
\includegraphics[width=0.95\columnwidth]{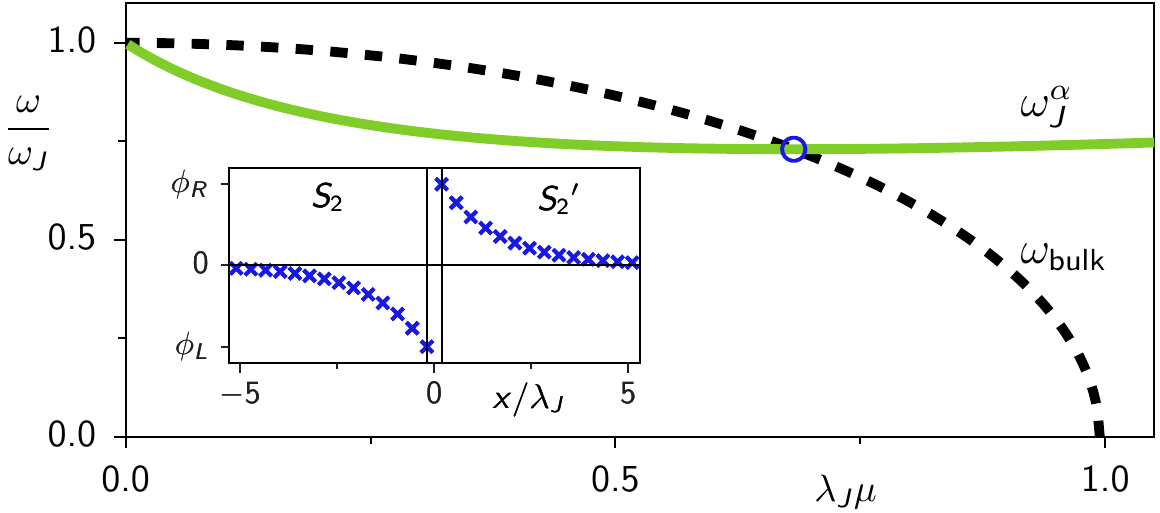}
\caption{
Bulk frequency $\omega_{\text{bulk}}$ from \Eq{eq:dispersion_discrete}
(black dashed) and plasma frequency $\omega_J^\alpha = \sqrt{2\pi I_c^\alpha /(\Phi_0 C_J^\alpha)}$ of the effective JJ, \Eqs{eq:Ceff}, \eqref{eq:Iceff} (green solid), 
as functions of inverse decay length $\mu$ of an LJJ edge state, cf. \Eq{eq:boundstate_rightLJJ}.
The edge states in LJJs $S_2$ and $S_2'$, with $\mu$ according to the intersection point of the two frequencies, are shown in the inset. 
Equations \eqref{eq:Ceff}, \eqref{eq:Iceff} are evaluated with interface parameters of IDSN gate, see caption of Fig.~\ref{fig:IDSN}.
}
 \label{fig:boundstatemodel}
\end{figure}

For concreteness we show here how the model is applied to the right LJJ $S_2'$; the same procedure applies to $S_2$.
The starting point for the analysis is the circuit Lagrangian of $S_2'$ together with the lower right interface JJ,
\begin{eqnarray}\label{eq:Lagr_rightLJJ}
&&\hspace*{-5mm}
\mathcal{L} = \left(\flqu\right)^2
 \left[ \frac{\CJab}{2} \dot{\phi}_0^2 + \sum_{n=1}^N \frac{C_J}{2} \left(\dot{\phi}_n\right)^2 
 \right] \nonumber \\
&& 
-\left(\flqu\right) \left[ \IJab (1-\cos\phi_0) + \sum_{n=1}^N I_c (1-\cos\phi_n) \right] 
\nonumber \\
 && -\left(\flqu\right)^2 \sum_{n=1}^N \frac{(\phi_n - \phi_{n-1})^2}{2 L}
\;.
\end{eqnarray}
For small amplitudes $\phi_n \ll \pi$, \Eq{eq:Lagr_rightLJJ} can be expanded to quadratic order in $\phi_n$.
We insert \Eq{eq:boundstate_rightLJJ} and extend the sums, $N\to\infty$,
under the assumption that the LJJ size is much larger than the decay length of the edge state,
$(N-1)a \gg \mu^{-1}$.
With these approximations the Lagrangian reduces to that of a simple LC-oscillator
of the edge phase $\phi_R$, 
\begin{eqnarray}\label{eq:Lagr_LCeff}
 \mathcal{L} 
 = \left(\flqu\right)^2 
 \left[\frac{C_J^{\alpha}}{2} \left(\dot{\phi}_R\right)^2
 - \frac{1}{2 L^\alpha} \phi_R^2\right]
\end{eqnarray}
with $\mu$-dependent effective parameters
\begin{eqnarray}
\label{eq:Ceff}
 C_J^{\alpha} &=& \CJab + C_J f(\mu) \\
\label{eq:invLeff}
 \frac{1}{L^{\alpha}} &=& \frac{2\pi}{\Phi_0} \left(\IJab + I_c f(\mu) \right) + \frac{g(\mu)}{L}
 \,.
\end{eqnarray}
Herein we have defined the functions
\begin{IEEEeqnarray}{lClCl}
 f(\mu) &=& \sum_{n=1}^\infty e^{-2\mu a n} 
     &=& (e^{2\mu a} - 1)^{-1} \\
 g(\mu) &=& \sum_{n=1}^\infty \left(e^{-\mu a n} - e^{-\mu a (n-1)} \right)^2
     &=& (e^{\mu a} - 1)^2 f(\mu) 
\end{IEEEeqnarray}
Since $\phi_R \ll \pi$ we can also interpret \Eq{eq:Lagr_LCeff} 
as the Lagrangian of a single JJ expanded to lowest orders. 
This JJ has (shunt) capacitance $C_J^{\alpha}(\mu)$
and critical current $I_c^{\alpha}(\mu)$, 
\begin{eqnarray}\label{eq:Iceff}
 I_c^{\alpha} = \flqu \frac{1}{L^{\alpha}(\mu)} 
 = \IJab + I_c f(\mu) + \flqu \frac{g(\mu)}{L} 
 \,.
\end{eqnarray}

By construction, the dynamics of this JJ should be equivalent to 
the dynamics of the lower right interface JJ together with 
the parallel LJJ $S_2'$, provided that it is only weakly excited by a fluxon from $S_1$.
We can apply the same model to the lower left interface JJ and LJJ $S_2$,
with edge state ansatz
\begin{equation}\label{eq:boundstate_leftLJJ}
 \phi_n = \phi_L(t) e^{-\mu a n} \,,
\end{equation}
where $x = -an \leq 0$ and $n=1,2,\ldots$ 
label the JJs in the LJJ with increasing distance from the interface.
Because of the left-right symmetry of the 2-bit gate structure, 
the resulting equivalent JJ of course has the same parameters, 
$C_J^{\alpha}(\mu)$ and $I_c^{\alpha}(\mu)$.

In the edge state ansatz, \Eqs{eq:boundstate_rightLJJ} and \eqref{eq:boundstate_leftLJJ}, 
the inverse decay length $\mu$ is still an unknown parameter and remains to be evaluated.
Here we fix $\mu$ by the condition that the plasma frequency 
$\omega_J^\alpha = \sqrt{2\pi I_c^\alpha /(\Phi_0 C_J^\alpha)}$ of the effective JJ
matches the frequency, 
with which the edge state oscillates, $\dot \phi_n = \omega \phi_n$.  
We approximate the latter as the oscillation frequency 
$\omega_{\text{bulk}}$ in the bulk of the LJJ, which follows 
from the bulk equations of motion,
\begin{equation}\label{eq:SGE_discrete}
 \ddot\phi_n - \frac{c^2}{a^2} \left(\phi_{n+1} - 2\phi_n + \phi_{n-1}\right) 
 + \omega_J^2 \sin\phi_n = 0
 \,.
\end{equation}
After linearizing for small excitations, $\phi_n \ll \pi$, 
and inserting \Eq{eq:boundstate_rightLJJ} or \Eq{eq:boundstate_leftLJJ} 
we obtain the bulk dispersion relation 
\begin{equation}\label{eq:dispersion_discrete}
 \omega_{\text{bulk}}^2 = \omega_J^2 + \frac{2 c^2}{a^2} \left(1 - \cosh(a \mu) \right)
 \,.
\end{equation}
In the homogeneous limit, $a \to 0$, where \Eq{eq:SGE_discrete} turns into the Sine-Gordon equation,
$\ddot\phi - c^2 \phi'' + \omega_J^2 \sin\phi = 0$,
\Eq{eq:dispersion_discrete} simplifies to $\omega^2 = \omega_J^2 (1 - \lambda_J^2 \mu^2)$.

\begin{figure}[bt]
 \includegraphics[width=0.98\columnwidth]{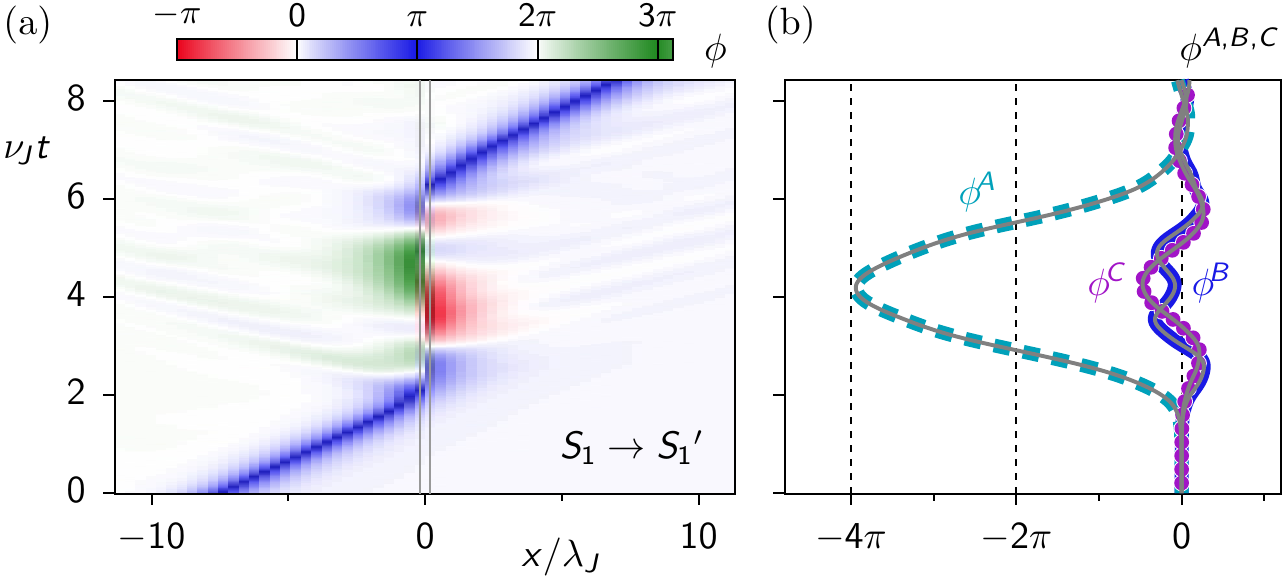}
 \caption{
 Dynamics of 1-bit gate circuit shown at bottom right of table 
 \ref{table:equivalences__interfaceAA1BB1CC1_Npar13_20200130}, with
 interface parameters stated ibid.
 By construction, the dynamics is approximately equivalent to the IDSN gate
 dynamics under single-fluxon input.
 This is evident when comparing the JJ-phases $\phi_n$ shown here in (a)
 with the JJ phases on LJJs $S_1$ and $S_1$ of the IDSN gate,
 shown in the first panel of Fig.~\ref{fig:IDSN}(b).
 Also the rail-JJ phases shown here in (b) for both the 1-bit interface 
 (thick lines in color) and the IDSN gate with single-fluxon input (thin gray lines)
 are basically indistinguishable on this scale.
 }
 \label{fig:IDSN_equiv_1bitgates}
\end{figure}

Fig.~\ref{fig:boundstatemodel} shows
$\omega_{\text{bulk}}$ from \Eq{eq:dispersion_discrete} (black dashed) 
together with the plasma frequency $\omega_J^\alpha$ of the effective JJ (green solid). 
For the latter we have used the parameters of the IDSN interface, cf.~the caption of Fig.~\ref{fig:IDSN}.
At the intersection point, $\lambda_J \mu = 0.68$, $\omega/\omega_J = 0.73$,
both relations between frequency $\omega$ and inverse decay length $\mu$ are satisfied. 
The inset of Fig.~\ref{fig:boundstatemodel} illustrates the edge states in LJJs $S_2$ and $S_2'$ for this $\mu$-value, and for $\phi_L=-\phi_R$.
With $\lambda_J \mu = 0.68$ and the parameters of the IDSN interface
we obtain $C_J^\alpha \approx 7.3 C_J$ and $I_c^\alpha \approx 3.9 I_c$. 
With these parameters the 1-bit structure shown in the right column of 
table \ref{table:equivalences__interfaceAA1BB1CC1_Npar13_20200130} 
is approximately 
dynamically equivalent to the 2-bit IDSN structure under single-fluxon input,
as demonstrated by Fig.~\ref{fig:IDSN_equiv_1bitgates},
in comparison with Fig.~\ref{fig:IDSN}(b).

\bibliographystyle{IEEEtran}

\vfill{} 
\begin{IEEEbiographynophoto}{Kevin D.~Osborn}
received a B.S.~in physics from Mary Washington University in Fredericksburg, Virginia and an M.S.~with thesis in physics from the University of Tennessee, Knoxville in 1992 and 1995, respectively. 
He then matriculated to the Physics Department at the University of Illinois at Urbana-Champaign and completed his Ph.D.~in 2001 with thesis studies of superfluid density fluctuations in high-temperature superconductors. 
Afterwards he transferred to the National Institute of Standards and Technology in Boulder, Colorado. 
Projects completed there included work on single-electron transistors over semiconductor quantum dots for quantum key distribution, and the development of superconducting qubits and coherent resonators.
In 2007, Dr.~Osborn started a research group in quantum computing at the Laboratory for Physical Sciences, with a joint appointment in the Physics Department, at the University of Maryland in College Park, MD. 
Most of Dr.~Osborn's over 40 publications are on qubits and two-level systems in materials, including custom quantum resonators to study them. 
In 2012, he hosted a workshop on reversible computing in Annapolis, MD, and since then he has kicked off research in that subfield of digital superconducting electronics. 
Dr.~Osborn is the first-listed inventor on two recent patents in reversible digital computing.
\end{IEEEbiographynophoto}
\vspace*{-30\baselineskip}
\begin{IEEEbiographynophoto}{W.~Wustmann}
received diploma and Ph.D.~degrees 
in theoretical physics at the Technical University in Dresden, Germany, in 2006 and 2010, respectively. 
Her graduate studies dealt with the statistical mechanics of periodically driven quantum systems.  
She then studied as a postdoc at Chalmers University in Gothenburg, Sweden, on the topic of parametric resonance in superconducting microwave circuits. 
Ever since this work she has contributed to and published with experimental groups.
In 2014 she joined the Laboratory for Physical Sciences in College Park, 
to work on theory and simulation of flux soliton dynamics in long Josephson junction 
circuits.
She is currently hosted by the University of Otago, New Zealand. 
\end{IEEEbiographynophoto}

\end{document}